\documentclass[traditabstract,longauth]{aa} % for the abstract without structuration 
\usepackage{graphicx,amsmath}
\usepackage{txfonts}
\usepackage{lscape}
\usepackage{color}
\usepackage{amssymb}
\usepackage{booktabs}
\usepackage{subfig}
\usepackage{url}                                                          %
\usepackage{hyperref}
\usepackage{natbib}
\bibpunct{(}{)}{;}{a}{}{,} % to follow the A&A style

\def\setsymbol#1#2{\expandafter\def\csname #1\endcsname{#2}}
\def\getsymbol#1{\csname #1\endcsname}

\def\Planck{\textit{Planck}}

\newbox\tablebox    \newdimen\tablewidth
\def\leaderfil{\leaders\hbox to 5pt{\hss.\hss}\hfil}
\def\endPlancktablewide{\tablewidth=\textwidth 
    $$\hss\copy\tablebox\hss$$
    \vskip-\lastskip\vskip -2pt}
\def\tablenote#1 #2\par{\begingroup \parindent=0.8em
    \abovedisplayshortskip=0pt\belowdisplayshortskip=0pt
    \noindent
    $$\hss\vbox{\hsize\tablewidth \hangindent=\parindent \hangafter=1 \noindent
    \hbox to \parindent{$^#1$\hss}\strut#2\strut\par}\hss$$
    \endgroup}
\def\doubleline{\vskip 3pt\hrule \vskip 1.5pt \hrule \vskip 5pt}

\def\L2{\ifmmode L_2\else $L_2$\fi}

\def\DeltaT{\ifmmode \Delta T\else $\Delta T$\fi}
\def\deltat{\ifmmode \Delta t\else $\Delta t$\fi}
\def\fknee{\ifmmode f_{\rm knee}\else $f_{\rm knee}$\fi}
\def\Fmax{\ifmmode F_{\rm max}\else $F_{\rm max}$\fi}
\def\solar{\ifmmode{\rm M}_{\mathord\odot}\else${\rm M}_{\mathord\odot}$\fi}
\def\Msolar{\ifmmode{\rm M}_{\mathord\odot}\else${\rm M}_{\mathord\odot}$\fi}
\def\Lsolar{\ifmmode{\rm L}_{\mathord\odot}\else${\rm L}_{\mathord\odot}$\fi}

\def\inv{\ifmmode^{-1}\else$^{-1}$\fi}
\def\mo{\ifmmode^{-1}\else$^{-1}$\fi}
\def\sup#1{\ifmmode ^{\rm #1}\else $^{\rm #1}$\fi}
\def\expo#1{\ifmmode \times 10^{#1}\else $\times 10^{#1}$\fi}
\def\,{\thinspace}
\def\lsim{\mathrel{\raise .4ex\hbox{\rlap{$<$}\lower 1.2ex\hbox{$\sim$}}}}
\def\gsim{\mathrel{\raise .4ex\hbox{\rlap{$>$}\lower 1.2ex\hbox{$\sim$}}}}

\def\simprop{\mathrel{\raise .4ex\hbox{\rlap{$\propto$}\lower 1.2ex\hbox{$\sim$}}}}
\def\deg{\ifmmode^\circ\else$^\circ$\fi}
\def\pdeg{\ifmmode $\setbox0=\hbox{$^{\circ}$}\rlap{\hskip.11\wd0 .}$^{\circ}
          \else \setbox0=\hbox{$^{\circ}$}\rlap{\hskip.11\wd0 .}$^{\circ}$\fi}
\def\arcs{\ifmmode {^{\scriptstyle\prime\prime}}
          \else $^{\scriptstyle\prime\prime}$\fi}
\def\arcm{\ifmmode {^{\scriptstyle\prime}}
          \else $^{\scriptstyle\prime}$\fi}
\newdimen\sa  \newdimen\sb
\def\parcs{\sa=.07em \sb=.03em
     \ifmmode \hbox{\rlap{.}}^{\scriptstyle\prime\kern -\sb\prime}\hbox{\kern -\sa}
     \else \rlap{.}$^{\scriptstyle\prime\kern -\sb\prime}$\kern -\sa\fi}
\def\parcm{\sa=.08em \sb=.03em
     \ifmmode \hbox{\rlap{.}\kern\sa}^{\scriptstyle\prime}\hbox{\kern-\sb}
     \else \rlap{.}\kern\sa$^{\scriptstyle\prime}$\kern-\sb\fi}
\def\ra[#1 #2 #3.#4]{#1\sup{h}#2\sup{m}#3\sup{s}\llap.#4}
\def\dec[#1 #2 #3.#4]{#1\deg#2\arcm#3\arcs\llap.#4}
\def\deco[#1 #2 #3]{#1\deg#2\arcm#3\arcs}
\def\rra[#1 #2]{#1\sup{h}#2\sup{m}}

\def\dots{\relax\ifmmode \ldots\else $\ldots$\fi}
\def\WHzsr{\ifmmode $W\,Hz\mo\,sr\mo$\else W\,Hz\mo\,sr\mo\fi}
\def\mHz{\ifmmode $\,mHz$\else \,mHz\fi}
\def\GHz{\ifmmode $\,GHz$\else \,GHz\fi}
\def\mKs{\ifmmode $\,mK\,s$^{1/2}\else \,mK\,s$^{1/2}$\fi}
\def\muKs{\ifmmode \,\mu$K\,s$^{1/2}\else \,$\mu$K\,s$^{1/2}$\fi}
\def\muKRJs{\ifmmode \,\mu$K$_{\rm RJ}$\,s$^{1/2}\else \,$\mu$K$_{\rm RJ}$\,s$^{1/2}$\fi}
\def\muKHz{\ifmmode \,\mu$K\,Hz$^{-1/2}\else \,$\mu$K\,Hz$^{-1/2}$\fi}
\def\MJysr{\ifmmode \,$MJy\,sr\mo$\else \,MJy\,sr\mo\fi}
\def\MJysrmK{\ifmmode \,$MJy\,sr\mo$\,mK$_{\rm CMB}\mo\else \,MJy\,sr\mo\,mK$_{\rm CMB}\mo$\fi}
\def\microns{\ifmmode \,\mu$m$\else \,$\mu$m\fi}

\def\muK{\ifmmode \,\mu$K$\else \,$\mu$\hbox{K}\fi}
\def\microK{\ifmmode \,\mu$K$\else \,$\mu$\hbox{K}\fi}
\def\muW{\ifmmode \,\mu$W$\else \,$\mu$\hbox{W}\fi}
\def\kms{\ifmmode $\,km\,s$^{-1}\else \,km\,s$^{-1}$\fi}
\def\kmsMpc{\ifmmode $\,\kms\,Mpc\mo$\else \,\kms\,Mpc\mo\fi}
\setsymbol{LFI:center:frequency:70GHz:units}{70.3\,GHz}
\setsymbol{LFI:center:frequency:44GHz:units}{44.1\,GHz}
\setsymbol{LFI:center:frequency:30GHz:units}{28.5\,GHz}

\setsymbol{LFI:center:frequency:70GHz}{70.3}
\setsymbol{LFI:center:frequency:44GHz}{44.1}
\setsymbol{LFI:center:frequency:30GHz}{28.5}

\setsymbol{LFI:center:frequency:LFI18:Rad:M:units}{71.7\GHz}
\setsymbol{LFI:center:frequency:LFI19:Rad:M:units}{67.5\GHz}
\setsymbol{LFI:center:frequency:LFI20:Rad:M:units}{69.2\GHz}
\setsymbol{LFI:center:frequency:LFI21:Rad:M:units}{70.4\GHz}
\setsymbol{LFI:center:frequency:LFI22:Rad:M:units}{71.5\GHz}
\setsymbol{LFI:center:frequency:LFI23:Rad:M:units}{70.8\GHz}
\setsymbol{LFI:center:frequency:LFI24:Rad:M:units}{44.4\GHz}
\setsymbol{LFI:center:frequency:LFI25:Rad:M:units}{44.0\GHz}
\setsymbol{LFI:center:frequency:LFI26:Rad:M:units}{43.9\GHz}
\setsymbol{LFI:center:frequency:LFI27:Rad:M:units}{28.3\GHz}
\setsymbol{LFI:center:frequency:LFI28:Rad:M:units}{28.8\GHz}
\setsymbol{LFI:center:frequency:LFI18:Rad:S:units}{70.1\GHz}
\setsymbol{LFI:center:frequency:LFI19:Rad:S:units}{69.6\GHz}
\setsymbol{LFI:center:frequency:LFI20:Rad:S:units}{69.5\GHz}
\setsymbol{LFI:center:frequency:LFI21:Rad:S:units}{69.5\GHz}
\setsymbol{LFI:center:frequency:LFI22:Rad:S:units}{72.8\GHz}
\setsymbol{LFI:center:frequency:LFI23:Rad:S:units}{71.3\GHz}
\setsymbol{LFI:center:frequency:LFI24:Rad:S:units}{44.1\GHz}
\setsymbol{LFI:center:frequency:LFI25:Rad:S:units}{44.1\GHz}
\setsymbol{LFI:center:frequency:LFI26:Rad:S:units}{44.1\GHz}
\setsymbol{LFI:center:frequency:LFI27:Rad:S:units}{28.5\GHz}
\setsymbol{LFI:center:frequency:LFI28:Rad:S:units}{28.2\GHz}

\setsymbol{LFI:center:frequency:LFI18:Rad:M}{71.7}
\setsymbol{LFI:center:frequency:LFI19:Rad:M}{67.5}
\setsymbol{LFI:center:frequency:LFI20:Rad:M}{69.2}
\setsymbol{LFI:center:frequency:LFI21:Rad:M}{70.4}
\setsymbol{LFI:center:frequency:LFI22:Rad:M}{71.5}
\setsymbol{LFI:center:frequency:LFI23:Rad:M}{70.8}
\setsymbol{LFI:center:frequency:LFI24:Rad:M}{44.4}
\setsymbol{LFI:center:frequency:LFI25:Rad:M}{44.0}
\setsymbol{LFI:center:frequency:LFI26:Rad:M}{43.9}
\setsymbol{LFI:center:frequency:LFI27:Rad:M}{28.3}
\setsymbol{LFI:center:frequency:LFI28:Rad:M}{28.8}
\setsymbol{LFI:center:frequency:LFI18:Rad:S}{70.1}
\setsymbol{LFI:center:frequency:LFI19:Rad:S}{69.6}
\setsymbol{LFI:center:frequency:LFI20:Rad:S}{69.5}
\setsymbol{LFI:center:frequency:LFI21:Rad:S}{69.5}
\setsymbol{LFI:center:frequency:LFI22:Rad:S}{72.8}
\setsymbol{LFI:center:frequency:LFI23:Rad:S}{71.3}
\setsymbol{LFI:center:frequency:LFI24:Rad:S}{44.1}
\setsymbol{LFI:center:frequency:LFI25:Rad:S}{44.1}
\setsymbol{LFI:center:frequency:LFI26:Rad:S}{44.1}
\setsymbol{LFI:center:frequency:LFI27:Rad:S}{28.5}
\setsymbol{LFI:center:frequency:LFI28:Rad:S}{28.2}

\setsymbol{LFI:white:noise:sensitivity:70GHz:units}{134.7\muKs}
\setsymbol{LFI:white:noise:sensitivity:44GHz:units}{164.7\muKs}
\setsymbol{LFI:white:noise:sensitivity:30GHz:units}{143.4\muKs}

\setsymbol{LFI:white:noise:sensitivity:70GHz}{134.7}
\setsymbol{LFI:white:noise:sensitivity:44GHz}{164.7}
\setsymbol{LFI:white:noise:sensitivity:30GHz}{143.4}

\setsymbol{LFI:white:noise:sensitivity:LFI18:Rad:M:units}{512.0\muKs}
\setsymbol{LFI:white:noise:sensitivity:LFI19:Rad:M:units}{581.4\muKs}
\setsymbol{LFI:white:noise:sensitivity:LFI20:Rad:M:units}{590.8\muKs}
\setsymbol{LFI:white:noise:sensitivity:LFI21:Rad:M:units}{455.2\muKs}
\setsymbol{LFI:white:noise:sensitivity:LFI22:Rad:M:units}{492.0\muKs}
\setsymbol{LFI:white:noise:sensitivity:LFI23:Rad:M:units}{507.7\muKs}
\setsymbol{LFI:white:noise:sensitivity:LFI24:Rad:M:units}{462.2\muKs}
\setsymbol{LFI:white:noise:sensitivity:LFI25:Rad:M:units}{413.6\muKs}
\setsymbol{LFI:white:noise:sensitivity:LFI26:Rad:M:units}{478.6\muKs}
\setsymbol{LFI:white:noise:sensitivity:LFI27:Rad:M:units}{277.7\muKs}
\setsymbol{LFI:white:noise:sensitivity:LFI28:Rad:M:units}{312.3\muKs}
\setsymbol{LFI:white:noise:sensitivity:LFI18:Rad:S:units}{465.7\muKs}
\setsymbol{LFI:white:noise:sensitivity:LFI19:Rad:S:units}{555.6\muKs}
\setsymbol{LFI:white:noise:sensitivity:LFI20:Rad:S:units}{623.2\muKs}
\setsymbol{LFI:white:noise:sensitivity:LFI21:Rad:S:units}{564.1\muKs}
\setsymbol{LFI:white:noise:sensitivity:LFI22:Rad:S:units}{534.4\muKs}
\setsymbol{LFI:white:noise:sensitivity:LFI23:Rad:S:units}{542.4\muKs}
\setsymbol{LFI:white:noise:sensitivity:LFI24:Rad:S:units}{399.2\muKs}
\setsymbol{LFI:white:noise:sensitivity:LFI25:Rad:S:units}{392.6\muKs}
\setsymbol{LFI:white:noise:sensitivity:LFI26:Rad:S:units}{418.6\muKs}
\setsymbol{LFI:white:noise:sensitivity:LFI27:Rad:S:units}{302.9\muKs}
\setsymbol{LFI:white:noise:sensitivity:LFI28:Rad:S:units}{285.3\muKs}

\setsymbol{LFI:white:noise:sensitivity:LFI18:Rad:M}{512.0}
\setsymbol{LFI:white:noise:sensitivity:LFI19:Rad:M}{581.4}
\setsymbol{LFI:white:noise:sensitivity:LFI20:Rad:M}{590.8}
\setsymbol{LFI:white:noise:sensitivity:LFI21:Rad:M}{455.2}
\setsymbol{LFI:white:noise:sensitivity:LFI22:Rad:M}{492.0}
\setsymbol{LFI:white:noise:sensitivity:LFI23:Rad:M}{507.7}
\setsymbol{LFI:white:noise:sensitivity:LFI24:Rad:M}{462.2}
\setsymbol{LFI:white:noise:sensitivity:LFI25:Rad:M}{413.6}
\setsymbol{LFI:white:noise:sensitivity:LFI26:Rad:M}{478.6}
\setsymbol{LFI:white:noise:sensitivity:LFI27:Rad:M}{277.7}
\setsymbol{LFI:white:noise:sensitivity:LFI28:Rad:M}{312.3}
\setsymbol{LFI:white:noise:sensitivity:LFI18:Rad:S}{465.7}
\setsymbol{LFI:white:noise:sensitivity:LFI19:Rad:S}{555.6}
\setsymbol{LFI:white:noise:sensitivity:LFI20:Rad:S}{623.2}
\setsymbol{LFI:white:noise:sensitivity:LFI21:Rad:S}{564.1}
\setsymbol{LFI:white:noise:sensitivity:LFI22:Rad:S}{534.4}
\setsymbol{LFI:white:noise:sensitivity:LFI23:Rad:S}{542.4}
\setsymbol{LFI:white:noise:sensitivity:LFI24:Rad:S}{399.2}
\setsymbol{LFI:white:noise:sensitivity:LFI25:Rad:S}{392.6}
\setsymbol{LFI:white:noise:sensitivity:LFI26:Rad:S}{418.6}
\setsymbol{LFI:white:noise:sensitivity:LFI27:Rad:S}{302.9}
\setsymbol{LFI:white:noise:sensitivity:LFI28:Rad:S}{285.3}

\setsymbol{LFI:knee:frequency:70GHz:units}{29.5\mHz}
\setsymbol{LFI:knee:frequency:44GHz:units}{56.2\mHz}
\setsymbol{LFI:knee:frequency:30GHz:units}{113.7\mHz}

\setsymbol{LFI:knee:frequency:70GHz}{29.5}
\setsymbol{LFI:knee:frequency:44GHz}{56.2}
\setsymbol{LFI:knee:frequency:30GHz}{113.7}

\setsymbol{LFI:knee:frequency:LFI18:Rad:M:units}{16.3\mHz}
\setsymbol{LFI:knee:frequency:LFI19:Rad:M:units}{15.1\mHz}
\setsymbol{LFI:knee:frequency:LFI20:Rad:M:units}{18.7\mHz}
\setsymbol{LFI:knee:frequency:LFI21:Rad:M:units}{37.2\mHz}
\setsymbol{LFI:knee:frequency:LFI22:Rad:M:units}{12.7\mHz}
\setsymbol{LFI:knee:frequency:LFI23:Rad:M:units}{34.6\mHz}
\setsymbol{LFI:knee:frequency:LFI24:Rad:M:units}{46.2\mHz}
\setsymbol{LFI:knee:frequency:LFI25:Rad:M:units}{24.9\mHz}
\setsymbol{LFI:knee:frequency:LFI26:Rad:M:units}{67.6\mHz}
\setsymbol{LFI:knee:frequency:LFI27:Rad:M:units}{187.4\mHz}
\setsymbol{LFI:knee:frequency:LFI28:Rad:M:units}{122.2\mHz}
\setsymbol{LFI:knee:frequency:LFI18:Rad:S:units}{17.7\mHz}
\setsymbol{LFI:knee:frequency:LFI19:Rad:S:units}{22.0\mHz}
\setsymbol{LFI:knee:frequency:LFI20:Rad:S:units}{8.7\mHz}
\setsymbol{LFI:knee:frequency:LFI21:Rad:S:units}{25.9\mHz}
\setsymbol{LFI:knee:frequency:LFI22:Rad:S:units}{15.8\mHz}
\setsymbol{LFI:knee:frequency:LFI23:Rad:S:units}{129.8\mHz}
\setsymbol{LFI:knee:frequency:LFI24:Rad:S:units}{100.9\mHz}
\setsymbol{LFI:knee:frequency:LFI25:Rad:S:units}{38.9\mHz}
\setsymbol{LFI:knee:frequency:LFI26:Rad:S:units}{58.9\mHz}
\setsymbol{LFI:knee:frequency:LFI27:Rad:S:units}{104.4\mHz}
\setsymbol{LFI:knee:frequency:LFI28:Rad:S:units}{40.7\mHz}

\setsymbol{LFI:knee:frequency:LFI18:Rad:M}{16.3}
\setsymbol{LFI:knee:frequency:LFI19:Rad:M}{15.1}
\setsymbol{LFI:knee:frequency:LFI20:Rad:M}{18.7}
\setsymbol{LFI:knee:frequency:LFI21:Rad:M}{37.2}
\setsymbol{LFI:knee:frequency:LFI22:Rad:M}{12.7}
\setsymbol{LFI:knee:frequency:LFI23:Rad:M}{34.6}
\setsymbol{LFI:knee:frequency:LFI24:Rad:M}{46.2}
\setsymbol{LFI:knee:frequency:LFI25:Rad:M}{24.9}
\setsymbol{LFI:knee:frequency:LFI26:Rad:M}{67.6}
\setsymbol{LFI:knee:frequency:LFI27:Rad:M}{187.4}
\setsymbol{LFI:knee:frequency:LFI28:Rad:M}{122.2}
\setsymbol{LFI:knee:frequency:LFI18:Rad:S}{17.7}
\setsymbol{LFI:knee:frequency:LFI19:Rad:S}{22.0}
\setsymbol{LFI:knee:frequency:LFI20:Rad:S}{8.7}
\setsymbol{LFI:knee:frequency:LFI21:Rad:S}{25.9}
\setsymbol{LFI:knee:frequency:LFI22:Rad:S}{15.8}
\setsymbol{LFI:knee:frequency:LFI23:Rad:S}{129.8}
\setsymbol{LFI:knee:frequency:LFI24:Rad:S}{100.9}
\setsymbol{LFI:knee:frequency:LFI25:Rad:S}{38.9}
\setsymbol{LFI:knee:frequency:LFI26:Rad:S}{58.9}
\setsymbol{LFI:knee:frequency:LFI27:Rad:S}{104.4}
\setsymbol{LFI:knee:frequency:LFI28:Rad:S}{40.7}

\setsymbol{LFI:slope:70GHz:units}{$-1.03$\mHz}
\setsymbol{LFI:slope:44GHz:units}{$-0.89$\mHz}
\setsymbol{LFI:slope:30GHz:units}{$-0.87$\mHz}

\setsymbol{LFI:slope:70GHz}{$-1.03$}
\setsymbol{LFI:slope:44GHz}{$-0.89$}
\setsymbol{LFI:slope:30GHz}{$-0.87$}

\setsymbol{LFI:slope:LFI18:Rad:M:units}{$-1.04$\mHz}
\setsymbol{LFI:slope:LFI19:Rad:M:units}{$-1.09$\mHz}
\setsymbol{LFI:slope:LFI20:Rad:M:units}{$-0.69$\mHz}
\setsymbol{LFI:slope:LFI21:Rad:M:units}{$-1.56$\mHz}
\setsymbol{LFI:slope:LFI22:Rad:M:units}{$-1.01$\mHz}
\setsymbol{LFI:slope:LFI23:Rad:M:units}{$-0.96$\mHz}
\setsymbol{LFI:slope:LFI24:Rad:M:units}{$-0.83$\mHz}
\setsymbol{LFI:slope:LFI25:Rad:M:units}{$-0.91$\mHz}
\setsymbol{LFI:slope:LFI26:Rad:M:units}{$-0.95$\mHz}
\setsymbol{LFI:slope:LFI27:Rad:M:units}{$-0.87$\mHz}
\setsymbol{LFI:slope:LFI28:Rad:M:units}{$-0.88$\mHz}
\setsymbol{LFI:slope:LFI18:Rad:S:units}{$-1.15$\mHz}
\setsymbol{LFI:slope:LFI19:Rad:S:units}{$-1.00$\mHz}
\setsymbol{LFI:slope:LFI20:Rad:S:units}{$-0.95$\mHz}
\setsymbol{LFI:slope:LFI21:Rad:S:units}{$-0.92$\mHz}
\setsymbol{LFI:slope:LFI22:Rad:S:units}{$-1.01$\mHz}
\setsymbol{LFI:slope:LFI23:Rad:S:units}{$-0.95$\mHz}
\setsymbol{LFI:slope:LFI24:Rad:S:units}{$-0.73$\mHz}
\setsymbol{LFI:slope:LFI25:Rad:S:units}{$-1.16$\mHz}
\setsymbol{LFI:slope:LFI26:Rad:S:units}{$-0.79$\mHz}
\setsymbol{LFI:slope:LFI27:Rad:S:units}{$-0.82$\mHz}
\setsymbol{LFI:slope:LFI28:Rad:S:units}{$-0.91$\mHz}

\setsymbol{LFI:slope:LFI18:Rad:M}{$-1.04$}
\setsymbol{LFI:slope:LFI19:Rad:M}{$-1.09$}
\setsymbol{LFI:slope:LFI20:Rad:M}{$-0.69$}
\setsymbol{LFI:slope:LFI21:Rad:M}{$-1.56$}
\setsymbol{LFI:slope:LFI22:Rad:M}{$-1.01$}
\setsymbol{LFI:slope:LFI23:Rad:M}{$-0.96$}
\setsymbol{LFI:slope:LFI24:Rad:M}{$-0.83$}
\setsymbol{LFI:slope:LFI25:Rad:M}{$-0.91$}
\setsymbol{LFI:slope:LFI26:Rad:M}{$-0.95$}
\setsymbol{LFI:slope:LFI27:Rad:M}{$-0.87$}
\setsymbol{LFI:slope:LFI28:Rad:M}{$-0.88$}
\setsymbol{LFI:slope:LFI18:Rad:S}{$-1.15$}
\setsymbol{LFI:slope:LFI19:Rad:S}{$-1.00$}
\setsymbol{LFI:slope:LFI20:Rad:S}{$-0.95$}
\setsymbol{LFI:slope:LFI21:Rad:S}{$-0.92$}
\setsymbol{LFI:slope:LFI22:Rad:S}{$-1.01$}
\setsymbol{LFI:slope:LFI23:Rad:S}{$-0.95$}
\setsymbol{LFI:slope:LFI24:Rad:S}{$-0.73$}
\setsymbol{LFI:slope:LFI25:Rad:S}{$-1.16$}
\setsymbol{LFI:slope:LFI26:Rad:S}{$-0.79$}
\setsymbol{LFI:slope:LFI27:Rad:S}{$-0.82$}
\setsymbol{LFI:slope:LFI28:Rad:S}{$-0.91$}

\setsymbol{LFI:FWHM:70GHz:units}{13\parcm01}
\setsymbol{LFI:FWHM:44GHz:units}{27\parcm92}
\setsymbol{LFI:FWHM:30GHz:units}{32\parcm65}

\setsymbol{LFI:FWHM:70GHz}{13.01}
\setsymbol{LFI:FWHM:44GHz}{27.92}
\setsymbol{LFI:FWHM:30GHz}{32.65}

\setsymbol{LFI:FWHM:LFI18:units}{13\parcm39}
\setsymbol{LFI:FWHM:LFI19:units}{13\parcm01}
\setsymbol{LFI:FWHM:LFI20:units}{12\parcm75}
\setsymbol{LFI:FWHM:LFI21:units}{12\parcm74}
\setsymbol{LFI:FWHM:LFI22:units}{12\parcm87}
\setsymbol{LFI:FWHM:LFI23:units}{13\parcm27}
\setsymbol{LFI:FWHM:LFI24:units}{22\parcm98}
\setsymbol{LFI:FWHM:LFI25:units}{30\parcm46}
\setsymbol{LFI:FWHM:LFI26:units}{30\parcm31}
\setsymbol{LFI:FWHM:LFI27:units}{32\parcm65}
\setsymbol{LFI:FWHM:LFI28:units}{32\parcm66}

\setsymbol{LFI:FWHM:LFI18}{13.39}
\setsymbol{LFI:FWHM:LFI19}{13.01}
\setsymbol{LFI:FWHM:LFI20}{12.75}
\setsymbol{LFI:FWHM:LFI21}{12.74}
\setsymbol{LFI:FWHM:LFI22}{12.87}
\setsymbol{LFI:FWHM:LFI23}{13.27}
\setsymbol{LFI:FWHM:LFI24}{22.98}
\setsymbol{LFI:FWHM:LFI25}{30.46}
\setsymbol{LFI:FWHM:LFI26}{30.31}
\setsymbol{LFI:FWHM:LFI27}{32.65}
\setsymbol{LFI:FWHM:LFI28}{32.66}

\setsymbol{LFI:FWHM:uncertainty:LFI18:units}{0.170\arcm}
\setsymbol{LFI:FWHM:uncertainty:LFI19:units}{0.174\arcm}
\setsymbol{LFI:FWHM:uncertainty:LFI20:units}{0.170\arcm}
\setsymbol{LFI:FWHM:uncertainty:LFI21:units}{0.156\arcm}
\setsymbol{LFI:FWHM:uncertainty:LFI22:units}{0.164\arcm}
\setsymbol{LFI:FWHM:uncertainty:LFI23:units}{0.171\arcm}
\setsymbol{LFI:FWHM:uncertainty:LFI24:units}{0.652\arcm}
\setsymbol{LFI:FWHM:uncertainty:LFI25:units}{1.075\arcm}
\setsymbol{LFI:FWHM:uncertainty:LFI26:units}{1.131\arcm}
\setsymbol{LFI:FWHM:uncertainty:LFI27:units}{1.266\arcm}
\setsymbol{LFI:FWHM:uncertainty:LFI28:units}{1.287\arcm}

\setsymbol{LFI:FWHM:uncertainty:LFI18}{0.170}
\setsymbol{LFI:FWHM:uncertainty:LFI19}{0.174}
\setsymbol{LFI:FWHM:uncertainty:LFI20}{0.170}
\setsymbol{LFI:FWHM:uncertainty:LFI21}{0.156}
\setsymbol{LFI:FWHM:uncertainty:LFI22}{0.164}
\setsymbol{LFI:FWHM:uncertainty:LFI23}{0.171}
\setsymbol{LFI:FWHM:uncertainty:LFI24}{0.652}
\setsymbol{LFI:FWHM:uncertainty:LFI25}{1.075}
\setsymbol{LFI:FWHM:uncertainty:LFI26}{1.131}
\setsymbol{LFI:FWHM:uncertainty:LFI27}{1.266}
\setsymbol{LFI:FWHM:uncertainty:LFI28}{1.287}

\setsymbol{HFI:center:frequency:100GHz:units}{100\,GHz}
\setsymbol{HFI:center:frequency:143GHz:units}{143\,GHz}
\setsymbol{HFI:center:frequency:217GHz:units}{217\,GHz}
\setsymbol{HFI:center:frequency:353GHz:units}{353\,GHz}
\setsymbol{HFI:center:frequency:545GHz:units}{545\,GHz}
\setsymbol{HFI:center:frequency:857GHz:units}{857\,GHz}

\setsymbol{HFI:center:frequency:100GHz}{100}
\setsymbol{HFI:center:frequency:143GHz}{143}
\setsymbol{HFI:center:frequency:217GHz}{217}
\setsymbol{HFI:center:frequency:353GHz}{353}
\setsymbol{HFI:center:frequency:545GHz}{545}
\setsymbol{HFI:center:frequency:857GHz}{857}

\setsymbol{HFI:Ndetectors:100GHz}{8}
\setsymbol{HFI:Ndetectors:143GHz}{11}
\setsymbol{HFI:Ndetectors:217GHz}{12}
\setsymbol{HFI:Ndetectors:353GHz}{12}
\setsymbol{HFI:Ndetectors:545GHz}{3}
\setsymbol{HFI:Ndetectors:857GHz}{4}

\setsymbol{HFI:FWHM:Maps:100GHz:units}{9\parcm88}
\setsymbol{HFI:FWHM:Maps:143GHz:units}{7\parcm18}
\setsymbol{HFI:FWHM:Maps:217GHz:units}{4\parcm87}
\setsymbol{HFI:FWHM:Maps:353GHz:units}{4\parcm65}
\setsymbol{HFI:FWHM:Maps:545GHz:units}{4\parcm72}
\setsymbol{HFI:FWHM:Maps:857GHz:units}{4\parcm39}
\setsymbol{HFI:FWHM:Maps:100GHz}{9.88}
\setsymbol{HFI:FWHM:Maps:143GHz}{7.18}
\setsymbol{HFI:FWHM:Maps:217GHz}{4.87}
\setsymbol{HFI:FWHM:Maps:353GHz}{4.65}
\setsymbol{HFI:FWHM:Maps:545GHz}{4.72}
\setsymbol{HFI:FWHM:Maps:857GHz}{4.39}

\setsymbol{HFI:beam:ellipticity:Maps:100GHz}{1.15}
\setsymbol{HFI:beam:ellipticity:Maps:143GHz}{1.01}
\setsymbol{HFI:beam:ellipticity:Maps:217GHz}{1.06}
\setsymbol{HFI:beam:ellipticity:Maps:353GHz}{1.05}
\setsymbol{HFI:beam:ellipticity:Maps:545GHz}{1.14}
\setsymbol{HFI:beam:ellipticity:Maps:857GHz}{1.19}

\setsymbol{HFI:FWHM:Mars:100GHz:units}{9\parcm37}
\setsymbol{HFI:FWHM:Mars:143GHz:units}{7\parcm04}
\setsymbol{HFI:FWHM:Mars:217GHz:units}{4\parcm68}
\setsymbol{HFI:FWHM:Mars:353GHz:units}{4\parcm43}
\setsymbol{HFI:FWHM:Mars:545GHz:units}{3\parcm80}
\setsymbol{HFI:FWHM:Mars:857GHz:units}{3\parcm67}

\setsymbol{HFI:FWHM:Mars:100GHz}{9.37}
\setsymbol{HFI:FWHM:Mars:143GHz}{7.04}
\setsymbol{HFI:FWHM:Mars:217GHz}{4.68}
\setsymbol{HFI:FWHM:Mars:353GHz}{4.43}
\setsymbol{HFI:FWHM:Mars:545GHz}{3.80}
\setsymbol{HFI:FWHM:Mars:857GHz}{3.67}

\setsymbol{HFI:beam:ellipticity:Mars:100GHz}{1.18}
\setsymbol{HFI:beam:ellipticity:Mars:143GHz}{1.03}
\setsymbol{HFI:beam:ellipticity:Mars:217GHz}{1.14}
\setsymbol{HFI:beam:ellipticity:Mars:353GHz}{1.09}
\setsymbol{HFI:beam:ellipticity:Mars:545GHz}{1.25}
\setsymbol{HFI:beam:ellipticity:Mars:857GHz}{1.03}

\setsymbol{HFI:CMB:relative:calibration:100GHz}{$\lsim 1\%$}
\setsymbol{HFI:CMB:relative:calibration:143GHz}{$\lsim 1\%$}
\setsymbol{HFI:CMB:relative:calibration:217GHz}{$\lsim 1\%$}
\setsymbol{HFI:CMB:relative:calibration:353GHz}{$\lsim 1\%$}
\setsymbol{HFI:CMB:relative:calibration:545GHz}{}
\setsymbol{HFI:CMB:relative:calibration:857GHz}{}

\setsymbol{HFI:CMB:absolute:calibration:100GHz}{$\lsim 2\%$}
\setsymbol{HFI:CMB:absolute:calibration:143GHz}{$\lsim 2\%$}
\setsymbol{HFI:CMB:absolute:calibration:217GHz}{$\lsim 2\%$}
\setsymbol{HFI:CMB:absolute:calibration:353GHz}{$\lsim 2\%$}
\setsymbol{HFI:CMB:absolute:calibration:545GHz}{}
\setsymbol{HFI:CMB:absolute:calibration:857GHz}{}

\setsymbol{HFI:FIRAS:gain:calibration:accuracy:statistical:100GHz}{}
\setsymbol{HFI:FIRAS:gain:calibration:accuracy:statistical:143GHz}{}
\setsymbol{HFI:FIRAS:gain:calibration:accuracy:statistical:217GHz}{}
\setsymbol{HFI:FIRAS:gain:calibration:accuracy:statistical:353GHz}{2.5\%}
\setsymbol{HFI:FIRAS:gain:calibration:accuracy:statistical:545GHz}{1\%}
\setsymbol{HFI:FIRAS:gain:calibration:accuracy:statistical:857GHz}{0.5\%}

\setsymbol{HFI:FIRAS:gain:calibration:accuracy:systematic:100GHz}{}
\setsymbol{HFI:FIRAS:gain:calibration:accuracy:systematic:143GHz}{}
\setsymbol{HFI:FIRAS:gain:calibration:accuracy:systematic:217GHz}{}
\setsymbol{HFI:FIRAS:gain:calibration:accuracy:systematic:353GHz}{}
\setsymbol{HFI:FIRAS:gain:calibration:accuracy:systematic:545GHz}{7\%}
\setsymbol{HFI:FIRAS:gain:calibration:accuracy:systematic:857GHz}{7\%}

\setsymbol{HFI:FIRAS:zero:point:accuracy:100GHz:units}{0.8\MJysr}
\setsymbol{HFI:FIRAS:zero:point:accuracy:143GHz:units}{}
\setsymbol{HFI:FIRAS:zero:point:accuracy:217GHz:units}{}
\setsymbol{HFI:FIRAS:zero:point:accuracy:353GHz:units}{1.4\MJysr}
\setsymbol{HFI:FIRAS:zero:point:accuracy:545GHz:units}{2.2\MJysr}
\setsymbol{HFI:FIRAS:zero:point:accuracy:857GHz:units}{1.7\MJysr}

\setsymbol{HFI:FIRAS:zero:point:accuracy:100GHz}{0.8}
\setsymbol{HFI:FIRAS:zero:point:accuracy:143GHz}{}
\setsymbol{HFI:FIRAS:zero:point:accuracy:217GHz}{}
\setsymbol{HFI:FIRAS:zero:point:accuracy:353GHz}{1.4}
\setsymbol{HFI:FIRAS:zero:point:accuracy:545GHz}{2.2}
\setsymbol{HFI:FIRAS:zero:point:accuracy:857GHz}{1.7}

\setsymbol{HFI:unit:conversion:100GHz:units}{0.2415\MJysrmK}
\setsymbol{HFI:unit:conversion:143GHz:units}{0.3694\MJysrmK}
\setsymbol{HFI:unit:conversion:217GHz:units}{0.4811\MJysrmK}
\setsymbol{HFI:unit:conversion:353GHz:units}{0.2883\MJysrmK}
\setsymbol{HFI:unit:conversion:545GHz:units}{0.05826\MJysrmK}
\setsymbol{HFI:unit:conversion:857GHz:units}{0.002238\MJysrmK}

\setsymbol{HFI:unit:conversion:100GHz}{0.2415}
\setsymbol{HFI:unit:conversion:143GHz}{0.3694}
\setsymbol{HFI:unit:conversion:217GHz}{0.4811}
\setsymbol{HFI:unit:conversion:353GHz}{0.2883}
\setsymbol{HFI:unit:conversion:545GHz}{0.05826}
\setsymbol{HFI:unit:conversion:857GHz}{0.002238}

\setsymbol{HFI:colour:correction:alpha=-2:V1.01:100GHz}{0.9893}
\setsymbol{HFI:colour:correction:alpha=-2:V1.01:143GHz}{0.9759}
\setsymbol{HFI:colour:correction:alpha=-2:V1.01:217GHz}{1.0007}
\setsymbol{HFI:colour:correction:alpha=-2:V1.01:353GHz}{1.0028}
\setsymbol{HFI:colour:correction:alpha=-2:V1.01:545GHz}{1.0019}
\setsymbol{HFI:colour:correction:alpha=-2:V1.01:857GHz}{0.9889}

\setsymbol{HFI:colour:correction:alpha=0:V1.01:100GHz}{1.0008}
\setsymbol{HFI:colour:correction:alpha=0:V1.01:143GHz}{1.0148}
\setsymbol{HFI:colour:correction:alpha=0:V1.01:217GHz}{0.9909}
\setsymbol{HFI:colour:correction:alpha=0:V1.01:353GHz}{0.9888}
\setsymbol{HFI:colour:correction:alpha=0:V1.01:545GHz}{0.9878}
\setsymbol{HFI:colour:correction:alpha=0:V1.01:857GHz}{1.0014}

\def\msol{{M$_{\odot}$}}
\def\xmm{XMM-{\it Newton} }

\def\xmm{{\it XMM-Newton}}
\def\planck{{\it Planck}}

\def \pn {\hbox{pn}} 
\def \mos {\hbox{\sc MOS}} 

\newfont{\gwpfont}{cmssq8 scaled 1000}

\newcommand {\apgt} {\ {\raise-.5ex\hbox{$\buildrel>\over\sim$}}\ }
\newcommand {\aplt} {\ {\raise-.5ex\hbox{$\buildrel<\over\sim$}}\ }
\def\M500{M_{500}}
\def\R500{R_{500}}
\def\Mgv{M_{\rm g,500}}

\def\YX {Y_{\rm X}}
\def\TX {T_{\rm X}}

\def \Rv {R_{500}}
\def\keV {\rm keV}
\def\Yv {Y_{500}}

\def\MYX {$M_{500}$--$Y_{\rm X}$}

\def\msol {{\rm M_{\odot}}}

\def\lesssim{\mathrel{\hbox{\rlap{\hbox{\lower4pt\hbox{$\sim$}}}\hbox{$<$}}}}
\def\gtrsim{\mathrel{\hbox{\rlap{\hbox{\lower4pt\hbox{$\sim$}}}\hbox{$>$}}}}

\def\p214 {PLCKG214.6+37.0}
\def\name {PLCKG214.6+37.0}

\newcommand{\propsim}{\lower 3pt \hbox{$\, \buildrel {\textstyle
       \propto}\over {\textstyle \sim}\,$}}

\begin{document}
%This author list corresponds to \title{Author list for PIP 16, Proj. Ref. 5.1/5.2/5.5: Planck intermediate results. VI. The dynamical structure of PLCKG214.6+37.0, a Planck-discovered supercluster}
%Prepared by R. Leonardi (rleonardi@sciops.esa.int), ESAC/ESA
%This version is from Wed Jul 11 15:42:59 2012 CET
%\subtitle{There are 202 co-authors in this list}
\author{\small
Planck Collaboration:
P.~A.~R.~Ade\inst{85}
\and
N.~Aghanim\inst{57}
\and
M.~Arnaud\inst{72}
\and
M.~Ashdown\inst{69, 5}
\and
F.~Atrio-Barandela\inst{18}
\and
J.~Aumont\inst{57}
\and
C.~Baccigalupi\inst{84}
\and
A.~Balbi\inst{35}
\and
A.~J.~Banday\inst{92, 8}
\and
R.~B.~Barreiro\inst{65}
% ...
}
\institute{\small
APC, AstroParticule et Cosmologie, Universit\'{e} Paris Diderot, CNRS/IN2P3, CEA/lrfu, Observatoire de Paris, Sorbonne Paris Cit\'{e}, 10, rue Alice Domon et L\'{e}onie Duquet, 75205 Paris Cedex 13, France\\
\and
Aalto University Mets\"{a}hovi Radio Observatory, Mets\"{a}hovintie 114, FIN-02540 Kylm\"{a}l\"{a}, Finland\\
\and
Agenzia Spaziale Italiana Science Data Center, c/o ESRIN, via Galileo Galilei, Frascati, Italy\\
\and
Agenzia Spaziale Italiana, Viale Liegi 26, Roma, Italy\\
\and
Astrophysics Group, Cavendish Laboratory, University of Cambridge, J J Thomson Avenue, Cambridge CB3 0HE, U.K.\\
\and
Atacama Large Millimeter/submillimeter Array, ALMA Santiago Central Offices, Alonso de Cordova 3107, Vitacura, Casilla 763 0355, Santiago, Chile\\
\and
CITA, University of Toronto, 60 St. George St., Toronto, ON M5S 3H8, Canada\\
\and
CNRS, IRAP, 9 Av. colonel Roche, BP 44346, F-31028 Toulouse cedex 4, France\\
\and
California Institute of Technology, Pasadena, California, U.S.A.\\
\and
Centre of Mathematics for Applications, University of Oslo, Blindern, Oslo, Norway\\
\and
Centro de Astrof\'{\i}sica, Universidade do Porto, Rua das Estrelas, 4150-762 Porto, Portugal\\
\and
Centro de Estudios de F\'{i}sica del Cosmos de Arag\'{o}n (CEFCA), Plaza San Juan, 1, planta 2, E-44001, Teruel, Spain\\
\and
Computational Cosmology Center, Lawrence Berkeley National Laboratory, Berkeley, California, U.S.A.\\
\and
Consejo Superior de Investigaciones Cient\'{\i}ficas (CSIC), Madrid, Spain\\
\and
DSM/Irfu/SPP, CEA-Saclay, F-91191 Gif-sur-Yvette Cedex, France\\
\and
DTU Space, National Space Institute, Juliane Mariesvej 30, Copenhagen, Denmark\\
\and
D\'{e}partement de Physique Th\'{e}orique, Universit\'{e} de Gen\`{e}ve, 24, Quai E. Ansermet,1211 Gen\`{e}ve 4, Switzerland\\
\and
Departamento de F\'{\i}sica Fundamental, Facultad de Ciencias, Universidad de Salamanca, 37008 Salamanca, Spain\\
\and
Departamento de F\'{\i}sica, Universidad de Oviedo, Avda. Calvo Sotelo s/n, Oviedo, Spain\\
\and
Department of Astrophysics, IMAPP, Radboud University, P.O. Box 9010, 6500 GL Nijmegen,  The Netherlands\\
\and
Department of Physics \& Astronomy, University of British Columbia, 6224 Agricultural Road, Vancouver, British Columbia, Canada\\
\and
Department of Physics and Astronomy, Dana and David Dornsife College of Letter, Arts and Sciences, University of Southern California, Los Angeles, CA 90089, U.S.A.\\
\and
Department of Physics, Gustaf H\"{a}llstr\"{o}min katu 2a, University of Helsinki, Helsinki, Finland\\
\and
Department of Physics, Princeton University, Princeton, New Jersey, U.S.A.\\
\and
Department of Physics, University of California, Berkeley, California, U.S.A.\\
\and
Department of Physics, University of California, One Shields Avenue, Davis, California, U.S.A.\\
\and
Department of Physics, University of California, Santa Barbara, California, U.S.A.\\
\and
Department of Physics, University of Illinois at Urbana-Champaign, 1110 West Green Street, Urbana, Illinois, U.S.A.\\
\and
Department of Statistics, Purdue University, 250 N. University Street, West Lafayette, Indiana, U.S.A.\\
\and
Dipartimento di Fisica e Astronomia G. Galilei, Universit\`{a} degli Studi di Padova, via Marzolo 8, 35131 Padova, Italy\\
\and
Dipartimento di Fisica, Universit\`{a} La Sapienza, P. le A. Moro 2, Roma, Italy\\
\and
Dipartimento di Fisica, Universit\`{a} degli Studi di Milano, Via Celoria, 16, Milano, Italy\\
\and
Dipartimento di Fisica, Universit\`{a} degli Studi di Trieste, via A. Valerio 2, Trieste, Italy\\
\and
Dipartimento di Fisica, Universit\`{a} di Ferrara, Via Saragat 1, 44122 Ferrara, Italy\\
\and
Dipartimento di Fisica, Universit\`{a} di Roma Tor Vergata, Via della Ricerca Scientifica, 1, Roma, Italy\\
\and
Dipartimento di Matematica, Universit\`{a} di Roma Tor Vergata, Via della Ricerca Scientifica, 1, Roma, Italy\\
\and
Discovery Center, Niels Bohr Institute, Blegdamsvej 17, Copenhagen, Denmark\\
\and
Dpto. Astrof\'{i}sica, Universidad de La Laguna (ULL), E-38206 La Laguna, Tenerife, Spain\\
\and
European Southern Observatory, ESO Vitacura, Alonso de Cordova 3107, Vitacura, Casilla 19001, Santiago, Chile\\
\and
European Space Agency, ESAC, Planck Science Office, Camino bajo del Castillo, s/n, Urbanizaci\'{o}n Villafranca del Castillo, Villanueva de la Ca\~{n}ada, Madrid, Spain\\
\and
European Space Agency, ESTEC, Keplerlaan 1, 2201 AZ Noordwijk, The Netherlands\\
\and
GEPI, Observatoire de Paris, Section de Meudon, 5 Place J. Janssen, 92195 Meudon Cedex, France\\
\and
Helsinki Institute of Physics, Gustaf H\"{a}llstr\"{o}min katu 2, University of Helsinki, Helsinki, Finland\\
\and
INAF - Osservatorio Astronomico di Padova, Vicolo dell'Osservatorio 5, Padova, Italy\\
\and
INAF - Osservatorio Astronomico di Roma, via di Frascati 33, Monte Porzio Catone, Italy\\
\and
INAF - Osservatorio Astronomico di Trieste, Via G.B. Tiepolo 11, Trieste, Italy\\
\and
INAF Istituto di Radioastronomia, Via P. Gobetti 101, 40129 Bologna, Italy\\
\and
INAF/IASF Bologna, Via Gobetti 101, Bologna, Italy\\
\and
INAF/IASF Milano, Via E. Bassini 15, Milano, Italy\\
\and
INFN, Sezione di Roma 1, Universit`{a} di Roma Sapienza, Piazzale Aldo Moro 2, 00185, Roma, Italy\\
\and
IPAG: Institut de Plan\'{e}tologie et d'Astrophysique de Grenoble, Universit\'{e} Joseph Fourier, Grenoble 1 / CNRS-INSU, UMR 5274, Grenoble, F-38041, France\\
\and
IUCAA, Post Bag 4, Ganeshkhind, Pune University Campus, Pune 411 007, India\\
\and
Imperial College London, Astrophysics group, Blackett Laboratory, Prince Consort Road, London, SW7 2AZ, U.K.\\
\and
Infrared Processing and Analysis Center, California Institute of Technology, Pasadena, CA 91125, U.S.A.\\
\and
Institut N\'{e}el, CNRS, Universit\'{e} Joseph Fourier Grenoble I, 25 rue des Martyrs, Grenoble, France\\
\and
Institut Universitaire de France, 103, bd Saint-Michel, 75005, Paris, France\\
\and
Institut d'Astrophysique Spatiale, CNRS (UMR8617) Universit\'{e} Paris-Sud 11, B\^{a}timent 121, Orsay, France\\
\and
Institut d'Astrophysique de Paris, CNRS (UMR7095), 98 bis Boulevard Arago, F-75014, Paris, France\\
\and
Institute for Space Sciences, Bucharest-Magurale, Romania\\
\and
Institute of Astro and Particle Physics, Technikerstrasse 25/8, University of Innsbruck, A-6020, Innsbruck, Austria\\
\and
Institute of Astronomy and Astrophysics, Academia Sinica, Taipei, Taiwan\\
\and
Institute of Astronomy, University of Cambridge, Madingley Road, Cambridge CB3 0HA, U.K.\\
\and
Institute of Theoretical Astrophysics, University of Oslo, Blindern, Oslo, Norway\\
\and
Instituto de Astrof\'{\i}sica de Canarias, C/V\'{\i}a L\'{a}ctea s/n, La Laguna, Tenerife, Spain\\
\and
Instituto de F\'{\i}sica de Cantabria (CSIC-Universidad de Cantabria), Avda. de los Castros s/n, Santander, Spain\\
\and
Jet Propulsion Laboratory, California Institute of Technology, 4800 Oak Grove Drive, Pasadena, California, U.S.A.\\
\and
Jodrell Bank Centre for Astrophysics, Alan Turing Building, School of Physics and Astronomy, The University of Manchester, Oxford Road, Manchester, M13 9PL, U.K.\\
\and
Kapteyn Astronomical Institute, University of Groningen, Landleven 12, 9747 AD Groningen, The Netherlands\\
\and
Kavli Institute for Cosmology Cambridge, Madingley Road, Cambridge, CB3 0HA, U.K.\\
\and
LAL, Universit\'{e} Paris-Sud, CNRS/IN2P3, Orsay, France\\
\and
LERMA, CNRS, Observatoire de Paris, 61 Avenue de l'Observatoire, Paris, France\\
\and
Laboratoire AIM, IRFU/Service d'Astrophysique - CEA/DSM - CNRS - Universit\'{e} Paris Diderot, B\^{a}t. 709, CEA-Saclay, F-91191 Gif-sur-Yvette Cedex, France\\
\and
Laboratoire Traitement et Communication de l'Information, CNRS (UMR 5141) and T\'{e}l\'{e}com ParisTech, 46 rue Barrault F-75634 Paris Cedex 13, France\\
\and
Laboratoire de Physique Subatomique et de Cosmologie, Universit\'{e} Joseph Fourier Grenoble I, CNRS/IN2P3, Institut National Polytechnique de Grenoble, 53 rue des Martyrs, 38026 Grenoble cedex, France\\
\and
Laboratoire de Physique Th\'{e}orique, Universit\'{e} Paris-Sud 11 \& CNRS, B\^{a}timent 210, 91405 Orsay, France\\
\and
Lawrence Berkeley National Laboratory, Berkeley, California, U.S.A.\\
\and
Max-Planck-Institut f\"{u}r Astrophysik, Karl-Schwarzschild-Str. 1, 85741 Garching, Germany\\
\and
Max-Planck-Institut f\"{u}r Extraterrestrische Physik, Giessenbachstra{\ss}e, 85748 Garching, Germany\\
\and
MilliLab, VTT Technical Research Centre of Finland, Tietotie 3, Espoo, Finland\\
\and
National University of Ireland, Department of Experimental Physics, Maynooth, Co. Kildare, Ireland\\
\and
Niels Bohr Institute, Blegdamsvej 17, Copenhagen, Denmark\\
\and
Observational Cosmology, Mail Stop 367-17, California Institute of Technology, Pasadena, CA, 91125, U.S.A.\\
\and
Optical Science Laboratory, University College London, Gower Street, London, U.K.\\
\and
SISSA, Astrophysics Sector, via Bonomea 265, 34136, Trieste, Italy\\
\and
School of Physics and Astronomy, Cardiff University, Queens Buildings, The Parade, Cardiff, CF24 3AA, U.K.\\
\and
Space Research Institute (IKI), Russian Academy of Sciences, Profsoyuznaya Str, 84/32, Moscow, 117997, Russia\\
\and
Space Sciences Laboratory, University of California, Berkeley, California, U.S.A.\\
\and
Stanford University, Dept of Physics, Varian Physics Bldg, 382 Via Pueblo Mall, Stanford, California, U.S.A.\\
\and
Tuorla Observatory, Department of Physics and Astronomy, University of Turku, V\"ais\"al\"antie 20, FIN-21500, Piikki\"o, Finland\\
\and
UPMC Univ Paris 06, UMR7095, 98 bis Boulevard Arago, F-75014, Paris, France\\
\and
Universit\'{e} Denis Diderot (Paris 7), 75205 Paris Cedex 13, France\\
\and
Universit\'{e} de Toulouse, UPS-OMP, IRAP, F-31028 Toulouse cedex 4, France\\
% ...
}

   \title{\Planck\ intermediate results. VI: The dynamical structure of
PLCKG214.6+37.0, a \Planck\ discovered
triple system of galaxy clusters.}
\titlerunning{The dynamical structure of
PLCKG214.6+37.0, a \Planck\ discovered
triple system of galaxy clusters}

%   \subtitle{I. Overviewing the $\kappa$-mechanism}

 %  \author{The Planck collaboration:Planck scientists and non Planck
 %  scientists members of WG5 }

%   \institute{Institute for Astronomy (IfA), University of Vienna,
%              T\"urkenschanzstrasse 17, A-1180 Vienna\\
%              \email{wuchterl@amok.ast.univie.ac.at}
%         \and
%             University of Alexandria, Department of Geography, ...\\
%             \email{c.ptolemy@hipparch.uheaven.space}
%             \thanks{The university of heaven temporarily does not
%                     accept e-mails}
%             }

   \date{Received ???; accepted ???}

% \abstract}{}{}{}{} 
% 5 {} token are mandatory
 
  \abstract
{The survey of galaxy clusters performed by \Planck\ through the
Sunyaev-Zeldovich effect has already discovered
many interesting objects, thanks to the whole coverage of the sky. One of the
SZ candidates detected in the early months of the mission near to the signal to
noise threshold, \name, was later
revealed by \xmm\ to be a triple system of galaxy clusters. We have
further investigated this puzzling system with a
multi-wavelength approach and we present here 
the results from  a deep \xmm\ re-observation. The characterisation of the 
physical properties of the three components has allowed us to build a template
model to extract the total SZ signal of this  system with \Planck\ data. We
partly reconciled the
discrepancy between the expected SZ signal from X-rays and the observed one,
which are now consistent at less than $1.2\,\sigma$. We measured the
redshift of the three components with the iron lines in the X-ray spectrum,
and confirmed that the three clumps are likely part of the same supercluster
structure. The analysis of the dynamical state of the three components, as well
as
the absence of detectable excess X-ray emission, suggest that we are witnessing
the formation of a massive cluster at an early phase of interaction.  

}


\authorrunning{Planck Collaboration}

 \maketitle
%
%________________________________________________________________
\section {Introduction}

\begin{figure*}[t]
\begin{centering}
\includegraphics[scale=1.,angle=0,keepaspectratio]{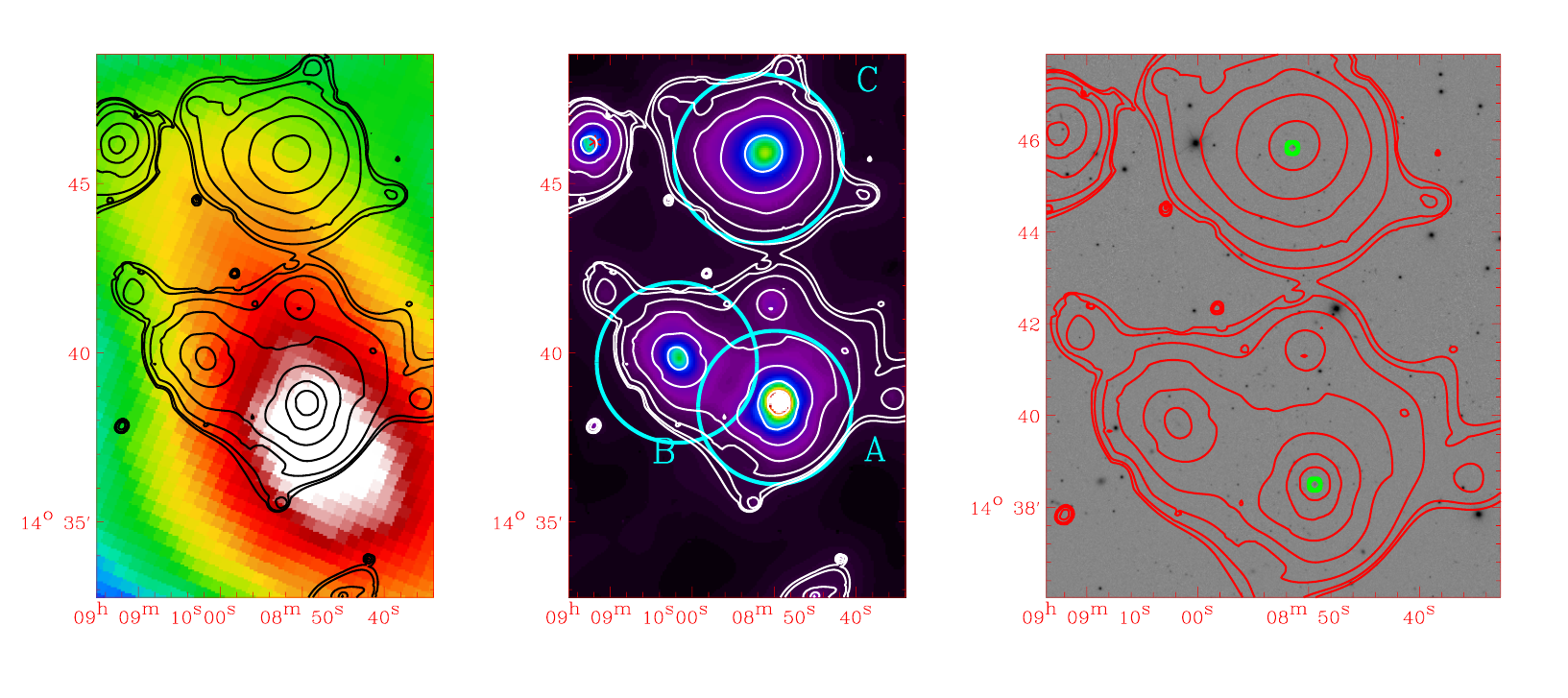}%
\end{centering}
\caption{The triple systems \name . \emph{Left panel:} 
\planck\
SZ reconstructed map
  (derived from the Modified Internal Linear Combination Analysis,
\citealt{hurier10}), oversampled and smoothed for displaying purposes. 
\emph{Middle panel:}
\xmm\ wavelet filtered image in the $[0.5$--$2.5]\,\keV$ (Sec 2.2).
The three components of \name\ are marked with circles (with radius
$\Rv$, Table 1)
and indicated with letters as in the
  text. The X-ray structure visible on the left (marked with *) is
    likely associated with the active galaxy
    SDSSJ$090912.15+144613.7$, with a spectroscopic redshift $z=0.767$.
\emph{Right panel:} SDSS $r-$band image of the \name\ field. In all panels,
North is up and East to the left and X-ray contours are overlaid.} 
\label{fig:sc}
\end{figure*}

Clusters of galaxies occupy a special position in the hierarchy of cosmic
structures: they are the largest objects that decoupled from the cosmic
expansion and that have had time to undergo gravitational collapse.  They are
thought to form via a hierarchical sequence of mergers and accretion of smaller
systems driven by gravity. During this process the intergalactic gas is heated
to high X-ray emitting temperatures by adiabatic compression and shocks and
settles in hydrostatic equilibrium within the cluster potential well.  
Sometimes galaxy clusters are found in multiple systems,
super-cluster structures which already decoupled from the Hubble flow and are
destined to collapse. The crowded environment of superclusters is an ideal
place to study the merging processes of individual components at an early
stage of merger and
witness the initial formation phase of very massive structures. Moreover, the
processes related to the contraction may increase the density of the
intercluster medium and make it observable with present instruments. An example
is the central complex of the Shapley concentration which has been the object
of extensive multi-wavelength observations with the aim of characterising the
merger processes in galaxy clusters \citep{kull99,
bardelli98a, rossetti06, giac05}.\\
Recently a new observational window has opened up for the study of the
astrophysics of galaxy clusters through the Sunyaev-Zeldovich effect
(\citealt{sun72}, SZ hereafter): a spectral distortion of the Cosmic Microwave
Background (CMB) generated through inverse Compton scattering of CMB
photons by thermal electrons in the intracluster medium (ICM). 
SZ surveys are discovering new clusters, some of which are
interesting merging systems (with ``El Gordo'', \citealt{men11}, being
probably the most spectacular example). \\
A key role in SZ science is now played by the \Planck
\footnote{\Planck\ (\url{http://www.esa.int/Planck}) is a project of 
the European Space Agency (ESA) with instruments provided by two 
scientific consortia funded by ESA member states (in particular the lead 
countries: France and Italy) with contributions from NASA (USA), and 
telescope reflectors provided in a collaboration between ESA and a 
scientific consortium led and funded by Denmark.} satellite.
Compared to other SZ surveys of galaxy clusters, \Planck\ has only moderate
band-dependent spatial resolution, but it possesses a unique nine band coverage
and more crucially it covers the whole sky. Therefore, it allows the detection
of the rarest objects: massive high-redshift systems \citep{planck2011-5.1c},
which are the most
sensitive to cosmology, and complex multiple systems, which are interesting for
the physics of structure formation.
Indeed, during the follow-up \xmm\ campaign of \Planck\ SZ candidates, we found
two new
double systems and two new triple systems of clusters
(\citealt{planck2011-5.1b},
hereafter Paper I). In all cases, the cumulative contribution predicted by
X-ray measurements was lower than the measured SZ signal, although 
compatible within three $\sigma$.\\
\name\ is the most massive and the X-ray brightest of the two \Planck\
discovered triple systems. The \xmm\ follow-up observations showed that the
\Planck\ SZ source candidate position is located $\sim 5\arcm$ from the two
southern components ($A$ and $B$).  A third subcomponent, $C$, lies
approximately $7\arcm$ to the North (Fig.~\ref{fig:sc}). The X-ray spectral
analysis of the component $A$ indicated a redshift of $z_{\rm
Fe}\sim~0.45$, consistent with two galaxies with spectroscopic redshift of $\sim
0.45$, close to the peaks of component $A$ and $C$. A cross-correlation with
SDSS-DR7 Luminous Red Galaxies and the Superclusters catalogue from the
SDSS-DR7 \citep{lii10} hinted that this triple system is
encompassed within a very large--scale structure located at $z\sim 0.45$, and
whose centroid lies about $2\deg$ to the South (see Appendix~B in Paper I for
further details).\\
In this paper we present new SZ measurements of
this object with \Planck\ and compare them with the results from a a deep \xmm\
re-observation. In
Sect.\, 2 we describe the analysis methods and the \Planck\ and \xmm\ data used
in this paper. In Sect.\, 3 we present our results obtained with X-ray and, in Sect.\, 4
compare them with available optical data from SDSS. In Sect.\,
5 we compare the X-ray results  with \Planck\
results.  In Sect.\, 6 we discuss our findings.\\
Throughout the paper we adopt a $\Lambda$CDM cosmology with $H_{0} = 70$~ km\
s$^{-1}$\ Mpc$^{-1}$, $\Omega_{\rm M} = 0.3$ and $\Omega_\Lambda= 0.7$.
At the nominal redshift of the supercluster, $z=0.45$, one arcminute
corresponds to 347 kpc.

\section{Observations}
\label{s:ob}

\subsection{Planck data and analysis}
\label{s:ob:planck}
\Planck\ \citep{tauber2010a,
Planck2011-1.1} is the third-generation space
mission to
 measure the anisotropy of the cosmic microwave background (CMB).  It observes
the sky in nine frequency bands covering 30--857\,GHz with high sensitivity and
angular resolution from 31\arcm\ to 5\arcm.  The Low Frequency Instrument (LFI;
\citealt{Mandolesi2010, Bersanelli2010, Planck2011-1.4}) covers the 30, 44, and
70\,GHz bands with amplifiers cooled to 20\,\hbox{K}.  The High Frequency
Instrument (HFI; \citealt{Lamarre2010, Planck2011-1.5}) covers the 100, 143,
217, 353, 545, and 857\,GHz bands with bolometers cooled to 0.1\,\hbox{K}. 
Polarisation is measured in all but the highest two bands \citep{Leahy2010,
Rosset2010}.  $A$ combination of radiative cooling and three mechanical coolers
produces the temperatures needed for the detectors and optics
\citep{Planck2011-1.3}.  Two data processing centres (DPCs) check and calibrate
the data and make maps of the sky \citep{Planck2011-1.7, Planck2011-1.6}. 
\Planck's sensitivity, angular resolution, and frequency coverage make it a
powerful instrument for galactic and extragalactic astrophysics as well as
cosmology.  Early astrophysics results are given in Planck Collaboration,
2011h--z.\\
Our results are based on the SZ signal as extracted from the six bands of
HFI corresponding to the nominal \Planck\
survey: 14 months, during which the whole sky was observed twice. We refer
to \citet{Planck2011-1.7} and \citet{Planck2011-1.6} for the generic scheme of
TOI processing and map making, as well as for the technical characteristics of
the maps used. We adopted a circular Gaussian as the beam pattern for each frequency
as described in \citet{Planck2011-1.7, Planck2011-1.6}.\\
The total SZ signal is characterised by the integrated Compton parameter $\Yv$
defined as $D_A^2(z)\Yv=(\sigma_T/m_ec^2)\int P_{th}dV$, where $D_A(z)$ is the
angular distance to a system at redshift $z$, 
$\sigma_T$ is the Thomson cross-section, $c$ the
speed of light, $m_e$ the rest mass of the electron, $P_{th}$ is the pressure of
thermal electrons and the integral is performed over a
sphere of radius $\R500$. \\
The extraction of the total SZ signal for this structure is more
complicated than for single clusters. Due  to its
 moderate spatial resolution, \Planck\ is not able to separate the contributions
of the three components from the whole signal. In Paper I, we estimated the total
flux assuming a single component with mass corresponding to
  the sum of the masses of the three clumps, following a universal pressure
profile \citep{arnaud10}
centred at the barycentre of the three components. This is obviously a simple
first-order approach. The wealth of information that is now available on this
system, has allowed us to build a more representative model. 
With the X-ray constraints on the structural properties of clumps $A$, $B$ and
$C$
(Sec. \ref{results_icm_global}), we are now able to build a three-component
model. As discussed in \citet{planck2011-5.1a}, our baseline
  pressure profile is the standard ``universal'' pressure profile
  derived by \citet{arnaud10}. We assumed this profile in each clump, parametrized in size by the
respective X-ray scale radius, $R_{500}$. The normalisations,
expressed
as integrated Comptonization parameters within $5R_{500}$ were tied up together
according to the ratio of their respective $Y_{X, 500}$, as determined from the
X-ray analysis.  Thus only one overall normalisation parameter remains to be
determined.
This template used under these assumptions and parametrisations together with
the
multi-frequency matched filter algorithm, MMF3 \citep{mel06}, directly
provides the integrated SZ flux over the whole super-structure.\\
 The two-dimensional reconstruction of the Comptonization parameter,
$y=(\sigma_T/m_ec^2)\int P_{th}dl$, provides a way to map the spatial
distribution of the thermal pressure integrated along the line of sight. 
This is performed using the modified internal linear combination
method (MILCA, \citealt{hurier10}) on the six
Planck 
all-sky maps from 100~GHz to 857~GHz. MILCA is a component separation approach aimed at 
extracting a chosen component (here the thermal SZ signal) from a 
multi-channel set of input maps. It is mainly based on the 
ILC approach \citep[e.g., ][]{eriksen04}, which searches for the linear 
combination of the input maps that minimises the variance of the final 
reconstructed map imposing spectral constraints. 

\subsection{\xmm\ observation and data reduction}
\label{sec:xmm_anal}
\name\ was re-observed by \xmm\ during AO10, for a nominal exposure
time of 65 ks. We produced calibrated event files from 
Observation Data Files (ODF) using v$.11.0$ of the \xmm\ Science Analysis
Software (SAS). We cleaned the event files from soft protons flares, using a
double filtering process (see \citealt{bour08} for details). After
the cleaning, the net exposure time is $\sim 47$ ks for the \mos\ detectors and
$\sim 37$ ks for the \pn. Since a quiescent component of soft protons may
survive the procedure described above, we calculated the ``in over out
ratio''
$R_{\textrm{SB}}$ \citep{del04} and we found values close to unity,
suggesting a negligible contribution of this component. We masked bright point
sources, detected in the \mos\ images as described in \citet{bour08}.
We performed the same data reduction procedure on the snapshot observation
0656200101, finding a cleaned exposure time of $\sim 15$ ks for the \mos\ and
$\sim 10$ for the pn.\\
We then combined the two observations and binned the photon events in
sky coordinates and energy cubes, matching the angular and spectral
resolution of each focal instrument. For spectroscopic and imaging
purposes, we associated an ``effective exposure'' and a ``background
noise'' cube to this photon cube (see \citealt{bour11} for details). 
The ``effective exposure'' is computed as a linear combination of CCD
exposure times related to individual observations, with local
corrections for useful CCD areas, RGS 
transmissions, and mirror vignetting factors. The ``background noise''
includes a set of particle background spectra modelled from
observations performed with the closed filter. 
Following an approach proposed in e.\,g.\,, \citet{lec08} or \citet{kuntz08}, this model sums a
quiescent continuum to a set of fluorescence emission lines convolved
with the energy response of each detector. Secondary background noise
components include the cosmic X-ray background and galactic
foregrounds. The cosmic X-ray background is modelled with an absorbed
power law of index $\Gamma = 1.42$ (e.\,g.\,, \citealt{lumb02}), while the
galactic foregrounds are modelled by the sum of two absorbed thermal
components accounting for the galactic transabsorption emission (
$kT_1 = 0.099$ keV, $kT_2 = 0.248$ keV, \citealt{kuntz00}). We
estimate emissivities of each of these components from a joint fit
of all background noise components in a region of the field of view
located beyond the supercluster boundary.\\
To estimate average ICM temperatures, $kT$
along the line of sight and for a given region of the field of view,
we added a source emission spectrum to the ``background noise'', and
fitted the spectral shape of the resulting function to the photon
energy distribution registered in the $0.3-10$ keV  energy band. In this
modelling, the source emission spectrum assumes a redshifted and $nH$
absorbed emission modelled from the Astrophysical Plasma Emission Code
(APEC, \citealt{smith01}), with the element abundances of \citet{grevesse98} and neutral hydrogen absorption cross sections of
\citet{balucinska}. It is corrected for effective
exposure, altered by the mirror effective areas, filter transmissions
and detector quantum efficiency, and convolved by a local energy
response matrix. \\
The X-ray image shown in Fig.\, \ref{fig:sc} is a wavelet filtered image, 
computed in the $0.5-2.5$ keV  energy
band. To generate this image, we corrected the photon map for
effective exposure and soft thresholded its undecimated B3--spline
wavelet coefficients \citep{starck07} to a $3\,\sigma$
level. In this procedure, significance thresholds have been directly
computed from the raw (Poisson distributed) photon map, following
the multi-scale variance stabilisation scheme introduced in \citet{zhang08}. We applied the same transformation to a ``background noise''
map, which we then subtracted from the image.

\section{Structure of the clusters from X-rays}

\subsection{Global analysis of the cluster components}
\label{results_icm_global}
Assuming that the three structures are located at the same redshift, $z = 0.45$
(the spectroscopic value found in Paper I), from the combination of the two
XMM-Newton observations we have
carried out an X-ray analysis for each component independently (we masked the
two other clumps while analysing the third one).
We extracted surface brightness
profiles of each component, centred on the X-ray peak, in the energy band
$0.5-2.5$ keV. We used the surface brightness profile to model the
three-dimensional density profile: the parametric density distribution
\citep{vik06} was projected, convolved with the PSF and fitted to the
observed surface brightness profile \citep{bour11}. From the density profile we measured the gas mass, $M_{g}$, that we
combined with the global temperature $\TX$, obtained with the spectral
analysis, to measure $\YX=M_{g}*\TX$ \citep{kra06}. We used the  \MYX\ scaling
relation in \citet{arnaud10} to estimate the total mass $\M500$, defined as the
mass corresponding to a density contrast $\delta=500$ with respect to the
critical density at the redshift of the cluster, $\rho_c(z)$, thus
$\M500=(4\pi/3)500\rho_c(z)\R500^3$.
The global
cluster parameters were estimated iteratively within $\R500$, until
convergence. \\
The resulting
global X-ray properties are summarised in Table \ref{table:x_prop}. The $\YX$
and $\M500$ values are slightly larger, but consistent within less than 
$1.5\,\sigma$ with the results shown in Paper I.  

\begin{table*}[tmb]                 % table* is a two-column table.  Drop the * for one column.
\begingroup
\newdimen\tblskip \tblskip=5pt
\caption{Physical properties of the components of \name\ .}              % title
\label{table:x_prop} 
\nointerlineskip
\vskip -3mm
\footnotesize
\setbox\tablebox=\vbox{
   \newdimen\digitwidth 
   \setbox0=\hbox{\rm 0} 
   \digitwidth=\wd0 
   \catcode`*=\active 
   \def*{\kern\digitwidth}
   \newdimen\signwidth 
   \setbox0=\hbox{+} 
   \signwidth=\wd0 
   \catcode`!=\active 
   \def!{\kern\signwidth}
\halign{  \hfil#\hfil& \hfil#\hfil\tabskip=0.5em & \hfil#\hfil\tabskip=0.5em& \hfil#\hfil\tabskip=0.5em&
  \hfil#\hfil\tabskip=0.5em& \hfil#\hfil\tabskip=0.5em& \hfil#\hfil\tabskip=0.5em& \hfil#\hfil\cr                       % Template goes here.
\noalign{\doubleline}
Component & $\rm{RA_X}$ & $\rm{Dec_X}$ & $\TX$ & $\Mgv$ & $\YX$ &
$\M500$ & $\R500$ \cr
  	      &  [hh:mm:ss] & [hh:mm:ss] & [keV] &  [$10^{14}\msol$] &
[$10^{14}\msol$ keV]   &  [$10^{14}\msol$] & [kpc] \cr                                   % Table headings go here.
\noalign{\vskip 3pt\hrule\vskip 5pt}
 A   & $09:08:49.6$  &  $+14:38:26.8$ & $3.6 \pm 0.4$ & 
$0.26 \pm 0.01$ &  $0.96 \pm 0.11$ & $2.22 \pm 0.16$ & $784 \pm 19$ \cr
B   & $09:09:01.8$   & $+14:39:45.6$ & $4.3 \pm 0.9$ &  $0.28 \pm
0.02$ & $ 1.2\pm 0.3$ & $2.5 \pm 0.4$ & $820 \pm 40$ \cr
C   & $09:08:51.2$  &  $+14:45:46.7$ &$5.3 \pm 0.9$
& $0.30 \pm 0.02$ & $1.6 \pm 0.3$ & $ 3.0 \pm 0.3$ & $864 \pm 33$ \cr	
                                   % Body of table goes here
\noalign{\vskip 5pt\hrule\vskip 3pt}}}
%\endPlancktable                    % ends one-column \halign
\endPlancktablewide                 % ends two-column \halign
%\tablenote a Footnote a.\par
%\tablenote b Footnote b.\par
\endgroup
\end{table*}

\subsection{Redshift estimates}
\begin{figure}
\begin{center}

\includegraphics[scale=1.,angle=0,keepaspectratio,
width=0.5\textwidth]{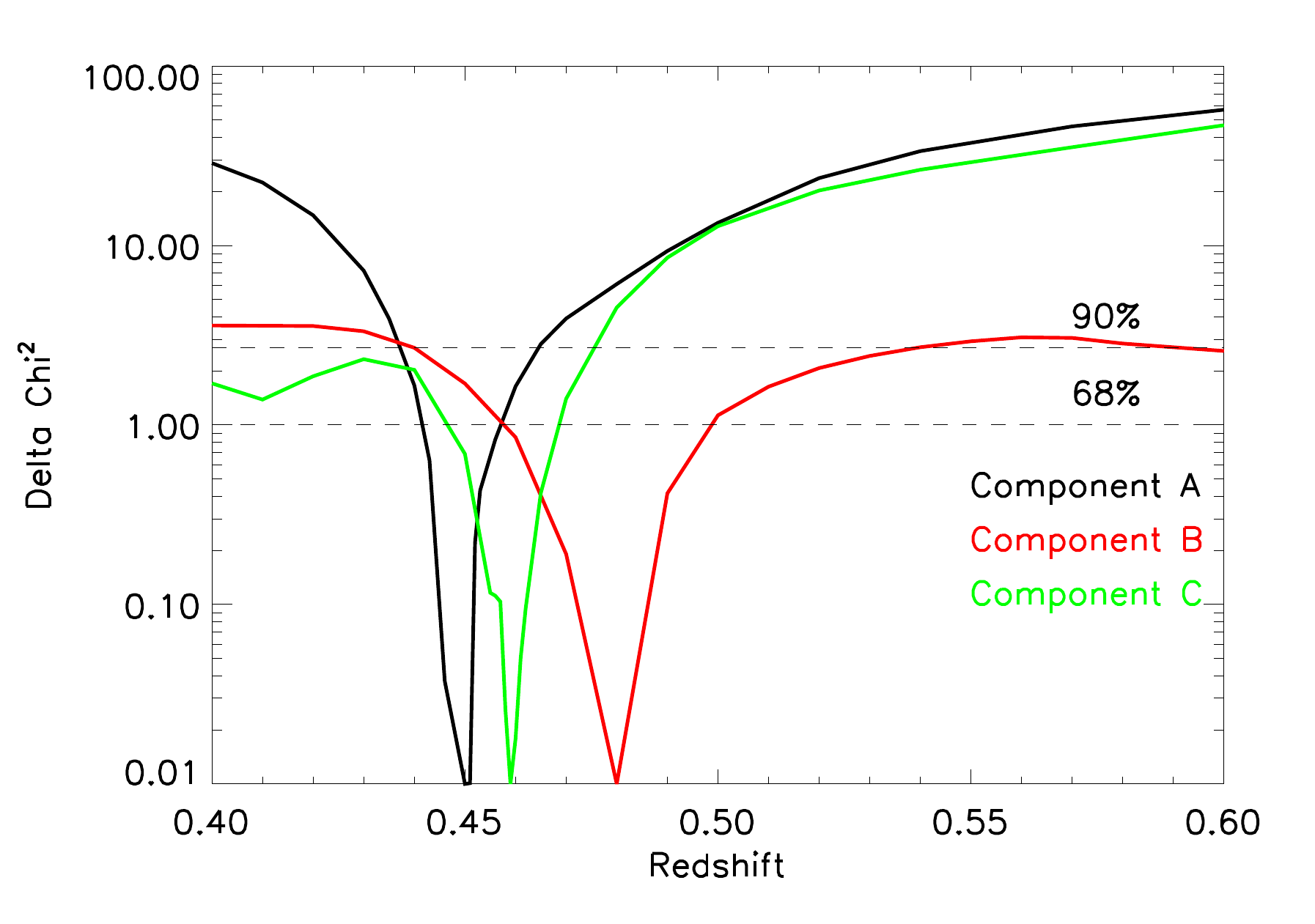}
\caption{Variation of $\chi^2$ when fitting the spectra for the redshift
measurement. Black, red, and
green lines represent component $A$, $B$, and $C$, respectively. Dashed thin
lines
correspond to the 68\% and 90 \% error range, respectively.}
\label{fig:deltachi}
\end{center}
\end{figure}

Crucial information on the nature of this triple system comes from the measurement of
the redshift of each component, allowing us to assess whether this is a
bound supercluster structure or a combination of unrelated objects
along the same line of sight.
In Paper I, a reliable redshift
measurement, obtained with the short \xmm\ observation, was available only for
component A. Its value ($z=0.45$) was consistent with the only two spectroscopic
redshifts available in this field and corresponding to the bright
central galaxies in components $A$ (SDSSJ$090849.38+143830.1$ $z=0.450$) and
$C$ (SDSSJ$090851.2+144551.0$ $z=0.452$). A photometric redshift, $z=0.46$,
was furthermore available for a bright galaxy (SDSSJ$090902.66+143948.1$) very
close to the peak of
component $B$. \\
With the new \xmm\ observation, we detect the iron K complex in each
clump. We extracted spectra in a circle centered on each component with
radius $\R500$ (Table \ref{table:x_prop}). 
We performed a more standard spectral extraction and analysis than the one
described in Sect.\, \ref{sec:xmm_anal}, extracting in each region the spectrum
for
each detector and its appropriate response (RMF) and ancillary (ARF) files.
We fitted spectra within XSPEC, modelling the instrumental and cosmic background
as in \citet{lec08}, leaving as free parameters of the fit the temperature,
metal abundance, redshift and normalisation of the cluster component (see
\citealt{planck2011-5.1b} for details). 
We first fitted spectra for each detector separately. While the MOS detectors
do not show any instrumental line in the whole $4-5$ keV range \citep{lec08},
the pn detector shows a faint fluorescent line\footnote{Ti $\rm{K}_\alpha$, 
\url{
http://xmm2.esac.esa.int/external/xmm_sw_cal/background/filter_closed/pn/mfreybe
rg-WA2-7.ps.gz}} in the spectral range where we expect
to find the redshifted cluster line. We
verified that this feature does not affect significantly our results, since
the pn
redshift and metal abundance are always consistent with at least one of the MOS
detectors. We report our results in Table \ref{table:xray_z}.\\
\begin{table*}[tmb]                 % table* is a two-column table.  Drop the * for one column.
\begingroup
\newdimen\tblskip \tblskip=5pt
\caption{Redshift measurements from the X-ray iron line for the components of
\name.}       % title
\label{table:xray_z}      % is used to refer this table in the text
\nointerlineskip
\vskip -3mm
\footnotesize
\setbox\tablebox=\vbox{
   \newdimen\digitwidth 
   \setbox0=\hbox{\rm 0} 
   \digitwidth=\wd0 
   \catcode`*=\active 
   \def*{\kern\digitwidth}
   \newdimen\signwidth 
   \setbox0=\hbox{+} 
   \signwidth=\wd0 
   \catcode`!=\active 
   \def!{\kern\signwidth}
\halign{      \hfil#\hfil& \hfil#\hfil\tabskip=0.5em &
  \hfil#\hfil\tabskip=0.5em& \hfil#\hfil\tabskip=0.5em& \hfil#\hfil\cr                      % Template goes here.
\noalign{\doubleline}
Component & MOS1 & MOS2 & pn & joint fit \cr                                    % Table headings go here.
\noalign{\vskip 3pt\hrule\vskip 5pt}
 A & $0.447\, (-0.013\; +0.024)$ & $0.446\,(-0.005\;+0.013)$ &
$0.441\,(-0.009\;+0.009)$ & $0.445\,(-0.006\;+0.006)$ (m1$+$m2$+$pn)\cr
B & $0.529\, (-0.018\; +0.024)$ & $0.472\,(-0.008\;+0.063)$ &
$0.475\,(-0.016\;+0.023)$ & $0.481\,(-0.011\;+0.013)$ (m2$+$pn) \cr
C & $0.469\, (-0.020 \;+0.020)$ & $0.434\,(-0.016\;+0.017)$ &
$0.463\,(-0.017\;+0.009)$ & $0.459\,(-0.010\;+0.010)$(m1$+$m2$+$pn) \cr
\noalign{\vskip 5pt\hrule\vskip 3pt}}}
%\endPlancktable                    % ends one-column \halign
\endPlancktablewide                 % ends two-column \halign
%\tablenote a Footnote a.\par
%\tablenote b Footnote b.\par
\endgroup
\end{table*}                        % table* is a two-column table.  Drop the * for one column.
For components  $A$ and $C$ the redshift measurements for
  each detector are consistent within one $\sigma$ and  we performed
joint fits combining all instruments (Table \ref{table:xray_z}).
For component $B$, the redshift measurement with MOS1 is larger than the
estimates with the other detectors, although consistent within two $
\sigma$. The joint fit of the three detectors in this case would lead
to a best fit value $z=0.516\,(-0.023,+0.014)$, while combining only
MOS2 and pn we find $0.481\,(-0.011,+0.013)$.
In the joint fit of the three detectors, the MOS1
spectrum drives the redshift estimate, leaving many residuals around
the position of the iron lines for the MOS2 and pn spectra. 
We performed simulations within XSPEC to quantify the probability
that, given the statistical quality of the spectra and the source to
background ratio, a redshift measurement as large as $z=0.529$, may
result just from statistical fluctuations of a spectrum with $z=0.48$. 
We assumed the best fit model of
the joint MOS2 and pn analysis as the input source and background
model with a redshift $z=0.48$ and 
we generated 1500 mock spectra that we fitted separately with
the same procedure we used for real spectra. 
We found a redshift as large as what we measured with the MOS1 in $3\%$ of
 the simulated spectra. With these simulations, we reproduced also the
 joint fit procedure: we performed 500 joint fits of three simulated
 spectra and found that a redshift as large as $0.516$ occurs with
 less than $1\%$ probability. 
Furthermore, we performed other simulations assuming the redshift
resulting from the joint fit of the three detectors $z=0.516$: the
probability of finding two redshifts as low as $z=0.475$ is about
$1.7\%$. The simulations we have described show that it is unlikely 
that the three redshift measurements and the joint fit  we obtained
for subcluster $B$
may result from statistical fluctuations of the same input spectrum,
suggesting a systematic origin for the discrepancy between MOS1 and
the other detectors. Indeed, the quality of the fit with the MOS1 data
alone is worse than with the other detectors, featuring many residuals
around the best fit model.  We verified the possibility of a
calibration issue affecting MOS1 by checking the position of the
instrumental lines: we did not find any significant systematic offset
for the bright low-energy Al and Si lines but the absence of strong
fluorescent lines between 2 and $5.4$ keV did not
allow us to test the calibration in the energy range we are interested
in.  Although the origin of the systematic difference of the MOS1
spectrum is still unclear, we decided to exclude this detector when
estimating the redshift of component $B$  (Table
\ref{table:xray_z}). Nonetheless, in the following, we will also
discuss the possibility that the cluster is located at the larger
redshift $z=0.516$. Concerning the components $A$ and $C$ the
redshift measurements are not significantly affected if we exclude the
MOS1 detector.\\
The redshift estimates we obtained from X-ray data for components $A$ ($z=0.445
\pm 0.006$) and $C$ ($z=0.46 \pm 0.01$) are nicely consistent with the
spectroscopic values found in the SDSS archive for their central brightest
galaxies ($0.450$ and $0.452$, respectively). 
Concerning component $B$, even  without considering the MOS1 detector, we still
find a larger best fit value ($z=0.48 \pm 0.01$) with respect to the other two
components. This is shown in Fig. \ref{fig:deltachi},
where we compare the
variation of $\chi^2$ for the joint fits (see Table \ref{table:xray_z}) for the
three components. While component $A$ and $C$ are consistent with being at the
same
redshift at less than one $\sigma$, component $B$ is likely located at a
larger redshift, although consistent at less than two $\sigma$ with the
position of the other clusters. Therefore, component $B$ is likely separated along the line
of sight from the two other components by $69\, (-30,\,+25)$ Mpc (150 Mpc, if we consider the redshift estimate obtained
with the three detectors). While this large separation suggests that
the cluster $B$ is not interacting with the other components, it is still consistent with the three objects being part of the same
supercluster structure \citep{bahcall99}.   \\ 
Using the best fit redshift estimates for the three
  components, we recomputed the physical parameters in Table 1.  While
the variations for cluster $A$ and $C$ are negligible, for cluster $B$
we found $\YX=(0.74\pm 0.19)\, 10^{14}\msol$ keV, $\M500=(1.89 \pm
0.20)\, 10^{14}\msol$ and $\R500=(726 \pm 25)$ kpc.

\subsection{Radial structural analysis of the components}

\begin{figure*}[t]
\begin{centering}
\includegraphics[scale=1.,angle=0,keepaspectratio,
width=\textwidth]{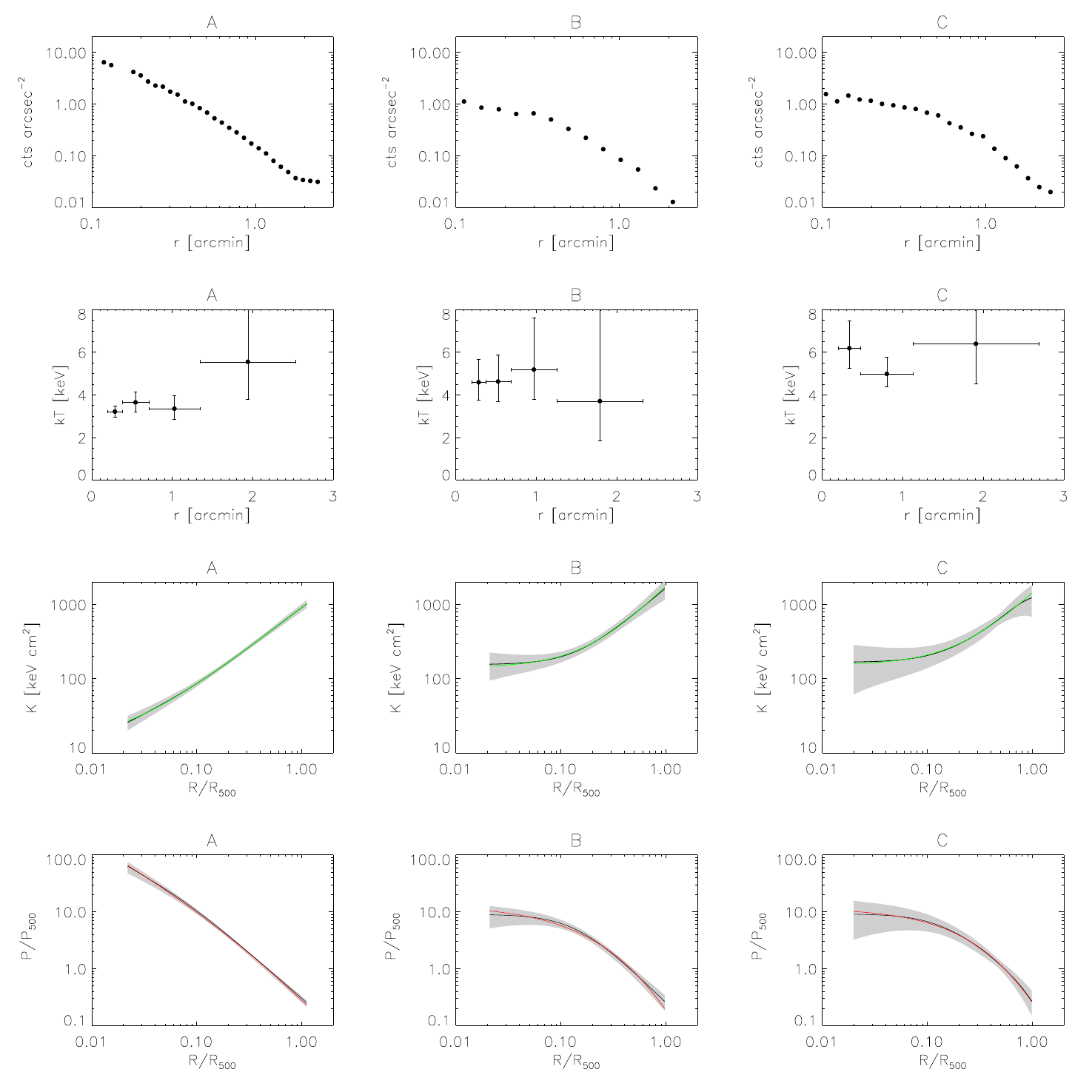}
\caption{Radial profiles for the relevant X-ray quantities for each component.
From top to bottom: surface brightness in the energy band $0.5-2.5$ keV,
projected temperature both as a function of the projected
  distance from the centre, three-dimensional entropy and pressure (rescaled by the
value at $\R500$) as a function of the distance from the
  centre in units of $\R500$ (with the values in Table 1). The black lines in the last two rows are the combination of
the density model with temperature to estimate entropy and pressure and the shaded
area shows the one $\sigma$ uncertainty. The red and green lines are our best
fit models, with the functions discussed in the text.}
\end{centering}
\end{figure*}
We performed a radial analysis of the X-ray observations of \name\ to study the
behaviour of the main thermo-dynamical quantities of the ICM. 
We extracted the surface brightness profile as discussed in Sec.
\ref{results_icm_global}: while the $A$ component shows a very peaked profile,
which might indicate a cool core state, the $B$ and $C$ components have flatter
profiles at the centre, a signature of an un-relaxed dynamical state. 
On more quantitative bases, we extracted the three-dimensional density profiles
for each component,
with the parametric procedure discussed in Sec.\,\ref{results_icm_global}, 
and computed the scaled central density, $n_{0}h(z)^{-2}$, where
$h(z)^2=\Omega_m(1+z)^3+\Omega_\Lambda$ is the ratio of the Hubble constant at
redshift $z$ with respect to its present value $H_0$. This parameter can
be used to classify clusters into cool-core ($n_{0}h(z)^{-2} >
4\times10^{-2}\,\rm{cm}^{-3}$) and non cool-core objects \citep{pra09}. As
expected, $A$ shows a central density  ($n_{0}h(z)^{-2} =
7\times10^{-2}\,\rm{cm}^{-3}$) typical of cool-core objects, while $B$ and $C$
show much lower central densities ($n_{0}h(z)^{-2} =
2\times10^{-3}\,\rm{cm}^{-3}$, both).\\
We extracted spectra in four (three for component $C$) annuli and fitted them
with a single-temperature absorbed model,
fixing as many components as we could because of the faintness of the source:
$ nH $ was fixed to the galactic value \citep{dic90}, redshift of
the
three components to $0.45$ and we fixed also all parameters of the background
components. In most cases we fixed also the metallicity to $0.3\, Z_{\odot}$,
except in the centre of subcluster $A$ where we could estimate an excess of
metal abundance ($Z=0.6 \pm 0.1\, Z_{\odot}$), as often found in cool
cores.
Due to the poor statistics, all temperature profiles are consistent at
one $\sigma$ with being flat and with the global values shown in Table
\ref{table:x_prop}, therefore from now on we
will consider them to be isothermal. \\
 We combined the three-dimensional density profile and the global
temperature to derive two other thermodynamic quantities: pressure and
entropy\footnote{The ``X-ray astronomer's entropy'' is defined as
$K=kT/n_e^{2/3}$, where $n_e$ is the electron density and $T$ the X-ray
temperature. This quantity is related to the thermodynamic entropy by a
logarithm and an additive constant.}.
Pressure is especially relevant to our analysis since it is the quantity that
is measured with the SZ effect. We have fitted the profiles with the 
model described in \citet{arnaud10}, the best fit parameters are consistent
with the ones for relaxed cool core objects for component $A$, and for disturbed
objects for component $B$ and $C$.\\
Entropy is a thermodynamic quantity that is connected both to the accretion
history of the cluster and to non gravitational processes.
If we fit
the profile with a power-law plus a constant, the central entropy $K_0$ is a
good indicator of the cool core state \citep{cava09}. The central entropy
values are essentially driven by the central densities because we assumed a
constant temperature, given the large uncertainties and poor resolution of the
temperature profiles. As expected, for subcluster $A$ we found $K_0=(13 \pm 2)\,
\rm{keV\,
cm^2}$, a central entropy typical
for cool core systems,
while for $B$ and $C$ we found
larger values
($K_0=142 \pm 10\, \rm{keV\,cm^2}$ and $K_0=153 \pm 18\, \rm{keV\,cm^2}$ respectively)
typical for unrelaxed objects.

\subsection{2D structure of the components and of the supercluster}
\begin{figure}
\begin{center}
\includegraphics[angle=0,keepaspectratio,
width=0.5\textwidth]{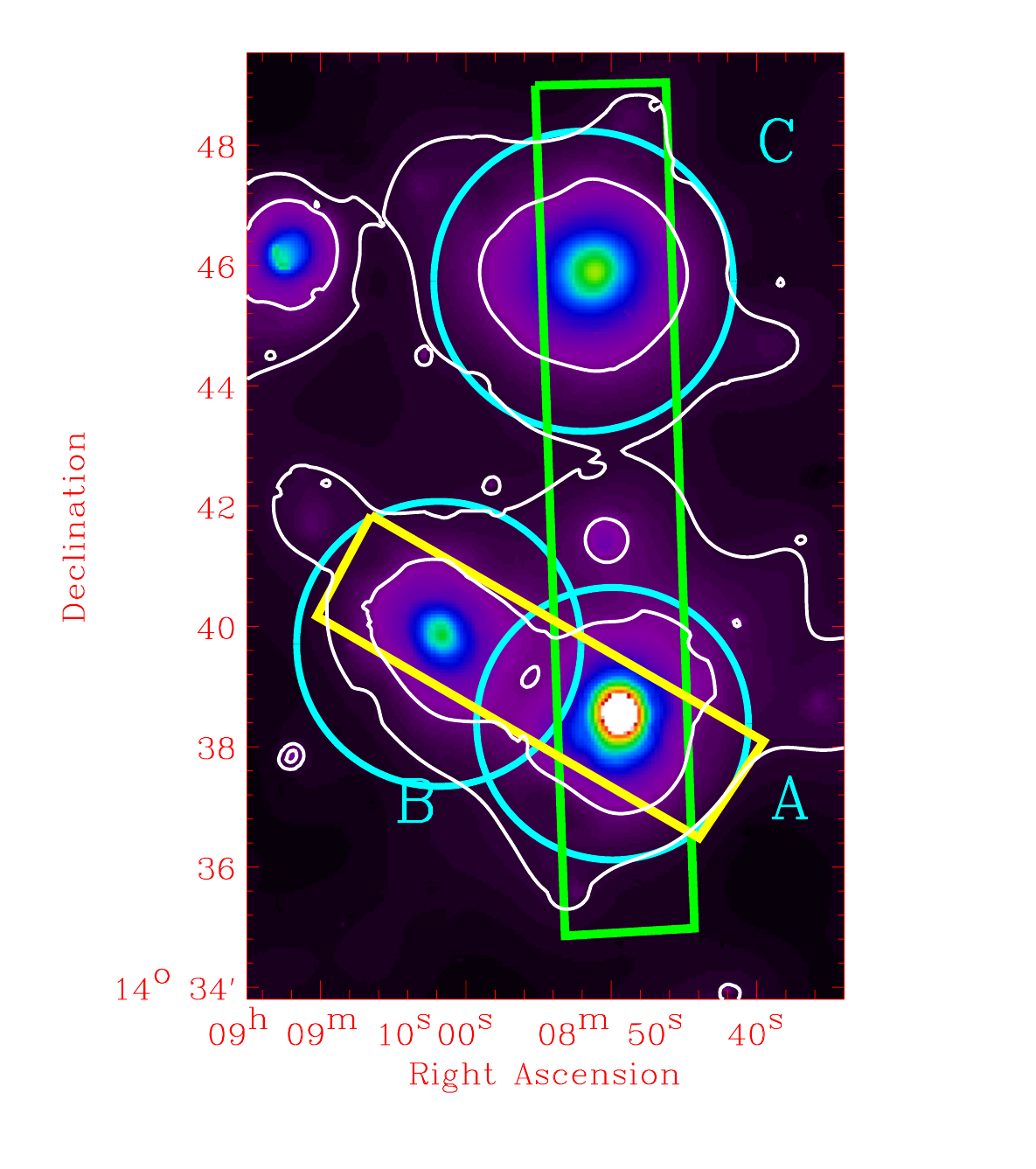}
\caption{\xmm\ wavelet filtered image of \name. Contours overlaid correspond to
the levels where we start to see connection between the components.
The green and yellow regions are the
ones where we extracted longitudinal profiles.}
\label{fig:lx_contours}
\end{center}
\end{figure}
\begin{figure*}
 \begin{centering}
\begin{minipage}[t]{0.90\textwidth}
\resizebox{\hsize}{!} {
\includegraphics[angle=0,keepaspectratio,scale=1]{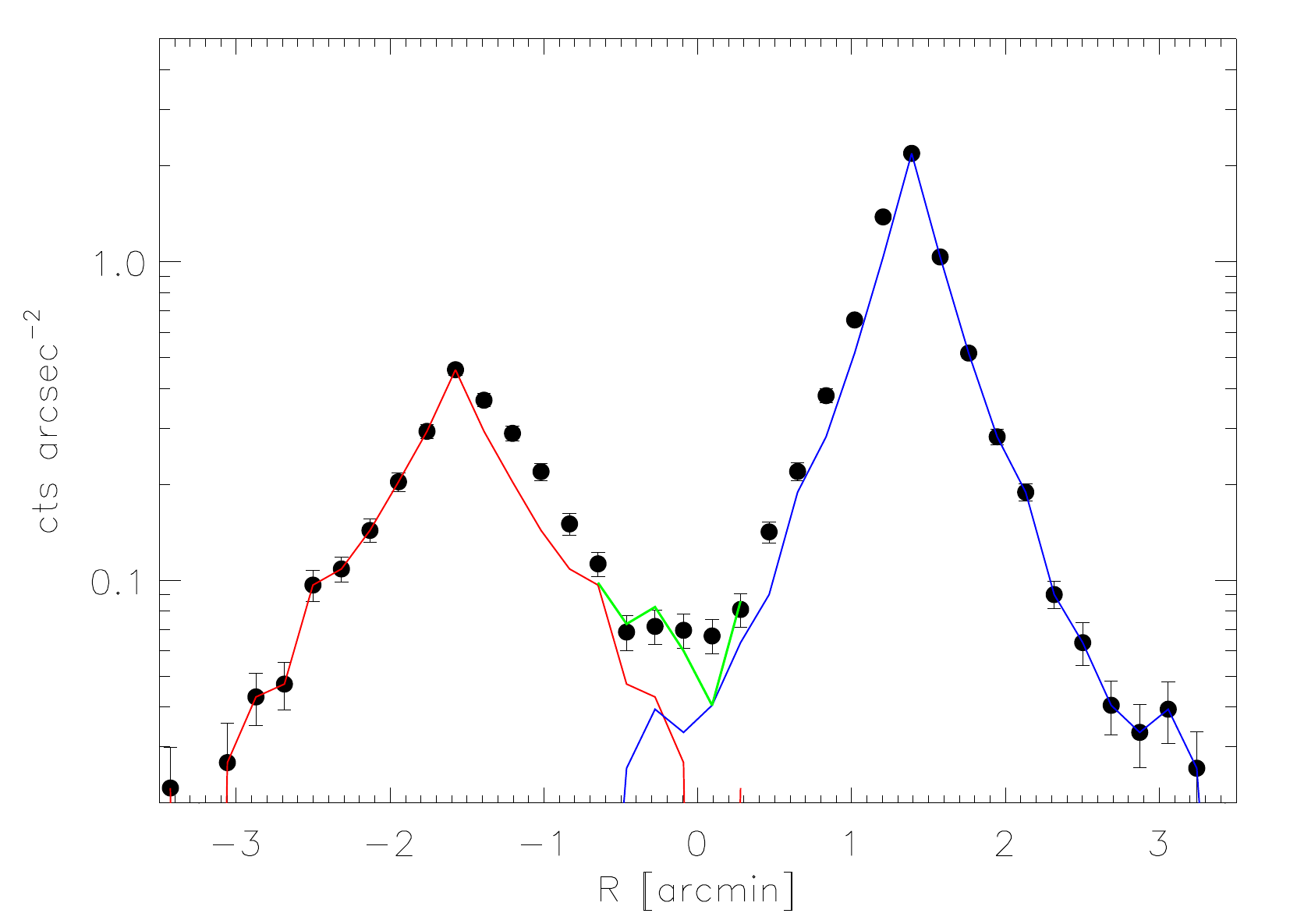}
\hspace{5mm}
\includegraphics[angle=0,keepaspectratio,scale=1]{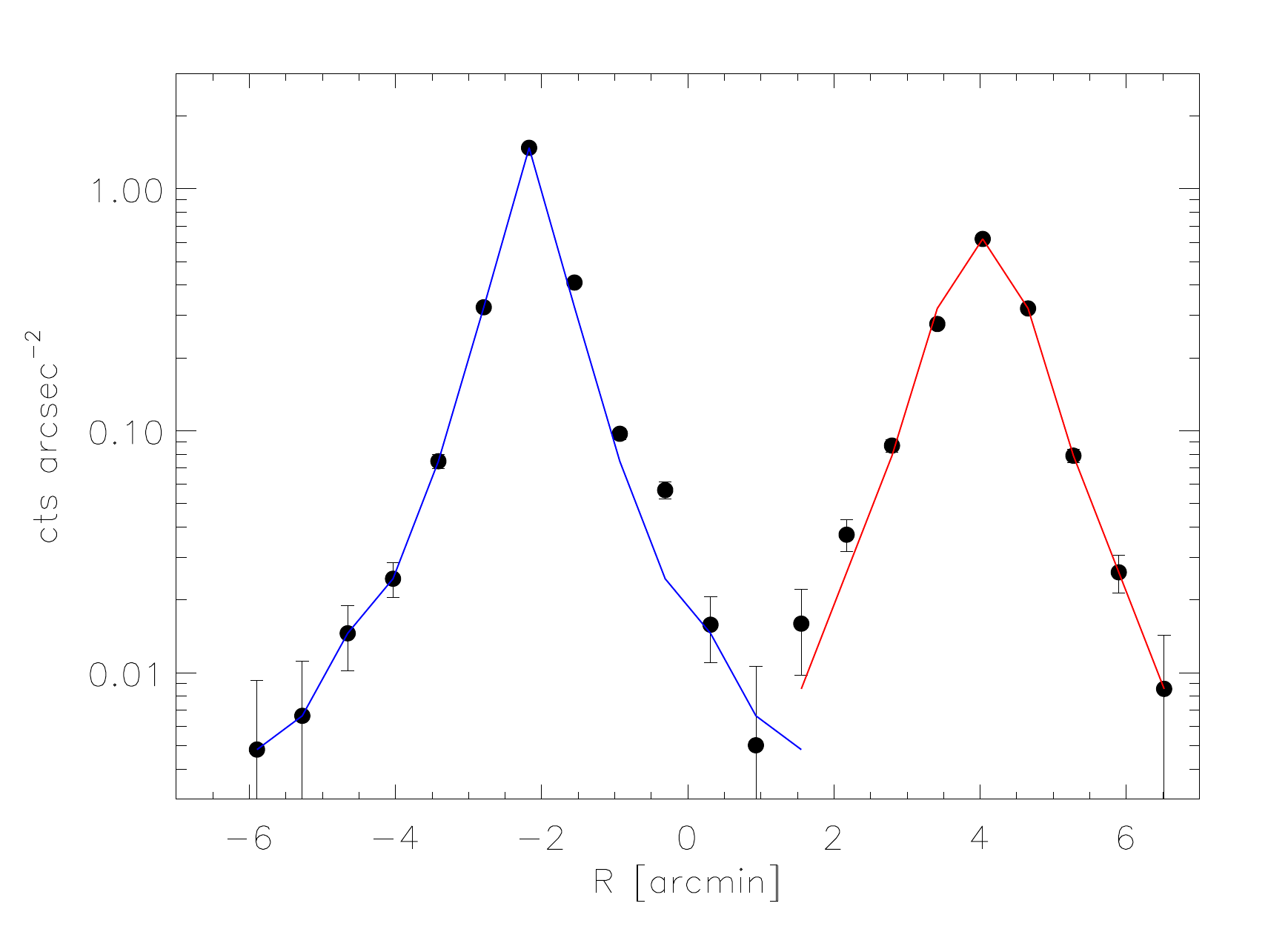}
}
\end{minipage}
%\label{fig:long_prof}
\caption{X-ray Longitudinal profile in the East-West (left panel) and
 North-South directions (right panel). {\it Left panel:} Negative distances
correspond to component $B$, positive ones to component $A$. The red and blue
lines
show the ``undisturbed'' models for components $B$ and $A$ (see text), while the
green line in the intersection region is the sum of the two models. 
{\it Right panel:} Negative
distances correspond to component $A$, positive ones to component $C$. The red
and blue lines show the ``undisturbed'' models for components $A$ and $C$ (see
text).}
\label{fig:long_prof}
\end{centering}
\end{figure*}

A qualitative analysis of the X-ray image (Fig.\, \ref{fig:lx_contours})
shows that the two southern components are apparently connected. Indeed the
X-ray surface brightness isophotes of component $B$ (Fig.\,\ref{fig:sc}) are
slightly elongated in the direction of component $A$, as often observed in
pairs of merging clusters (e.\,g.\,, the three systems in
\citealt{maurogordato11} and the pair A399-A401 \citealt{sake04}). 
We investigated visually the possible connection
between the components by  drawing constant surface brightness
contours in the X-ray image. The appearance of the contours may provide
information about the two-dimensional distribution of the intracluster
gas and the possible contamination by residual point sources.
A connection between the
components $A$ and $B$ is robustly detected, at a contour level above the
background intensity  (inner contour in Fig.\,
\ref{fig:lx_contours}). 
However, with this simple analysis it is not possible to assess whether this
connection is real or only a projection effect.
We used the same method between $A$ and $C$ where
we start to see a connection at a much lower intensity,
about 25$\%$ of the
level of the background model in the same region (outer contour in Fig.\,
\ref{fig:lx_contours}). In this regime, it might still be possible that the
connection between the two components is due to uncertainties in background
estimation or to residual point sources.\\
On more quantitative bases, we extracted longitudinal surface
brightness profiles in the East-West direction across components
$A$ and $B$,  and in the North-South direction,
across components $A$ and $C$ (cyan and green boxes in
Fig.\,\ref{fig:lx_contours}).
The profile
across components $A$ and $B$ (Fig.\,\ref{fig:long_prof}, left panel) shows
clearly enhanced emission with respect to the opposite direction between the two
clumps: we
modelled the emission of each component by taking the data in the external part
of the pair and  we project it symmetrically in the direction of the possible
interaction (blue and red lines). In the region where the two emissions overlap,
we summed the two models and found their sum to be consistent with the data.
The two objects are very close in the plane of the sky and their emissions
apparently overlap at less than $\R500$ (Fig.\,\ref{fig:sc}); if they were
located at the same distance from us and interacting  we would expect
to see compression and enhanced X-ray emission between the two objects. This is
not the case here, so our results argue in favour of a separation along the
line of sight of the two components, possibly still in an early
phase of
interaction.\\
Concerning components $A$ and $C$, their distance in the plane of the sky is
$7.4\arcmin$, corresponding to $\sim 2.5$ Mpc at $z=0.45$. 
The analysis of the longitudinal surface brightness profile across them
(Fig.\,\ref{fig:long_prof}, right panel) confirms our earlier indication:
the emission in the intersection region is not significantly detected and is
consistent with the ``undisturbed'' model (derived as before). \\
These results suggest that the three clusters, while likely belonging to the
same structure, have not started to interact yet.

\section{Comparison with optical data}
Since the sky region of \name\ is covered by the Sloan Digital Sky
Survey\footnote{\url{http://www.sdss.org/}}, we retrieved 
a galaxy catalogue from SDSS Data Release 8 (DR8). It covers a
circular
area of 20$\arcm$ radius around the barycentre of  \name\ and
includes optical magnitudes and photometric redshifts (see
\citealt{sdss_dr7} for description of measurements and calibrations of
photometric readshifts in SDSS DR8). It contains about 2000
objects, $\sim 900$ of which are in the redshift range $0.35-0.6$.
Unfortunately, spectroscopic redshifts are available only for the brightest
central galaxies of components $A$ and $C$, thus we relied on photometric
redshifts alone in our analysis. Spectroscopic information on this system
will be available from our follow-up program and will be discussed in a
forthcoming paper.
 
\subsection{Photometric redshifts of the three components}
\label{sec:photoz}
\begin{figure}
\begin{center}
\includegraphics[keepaspectratio,
width=0.5\textwidth]{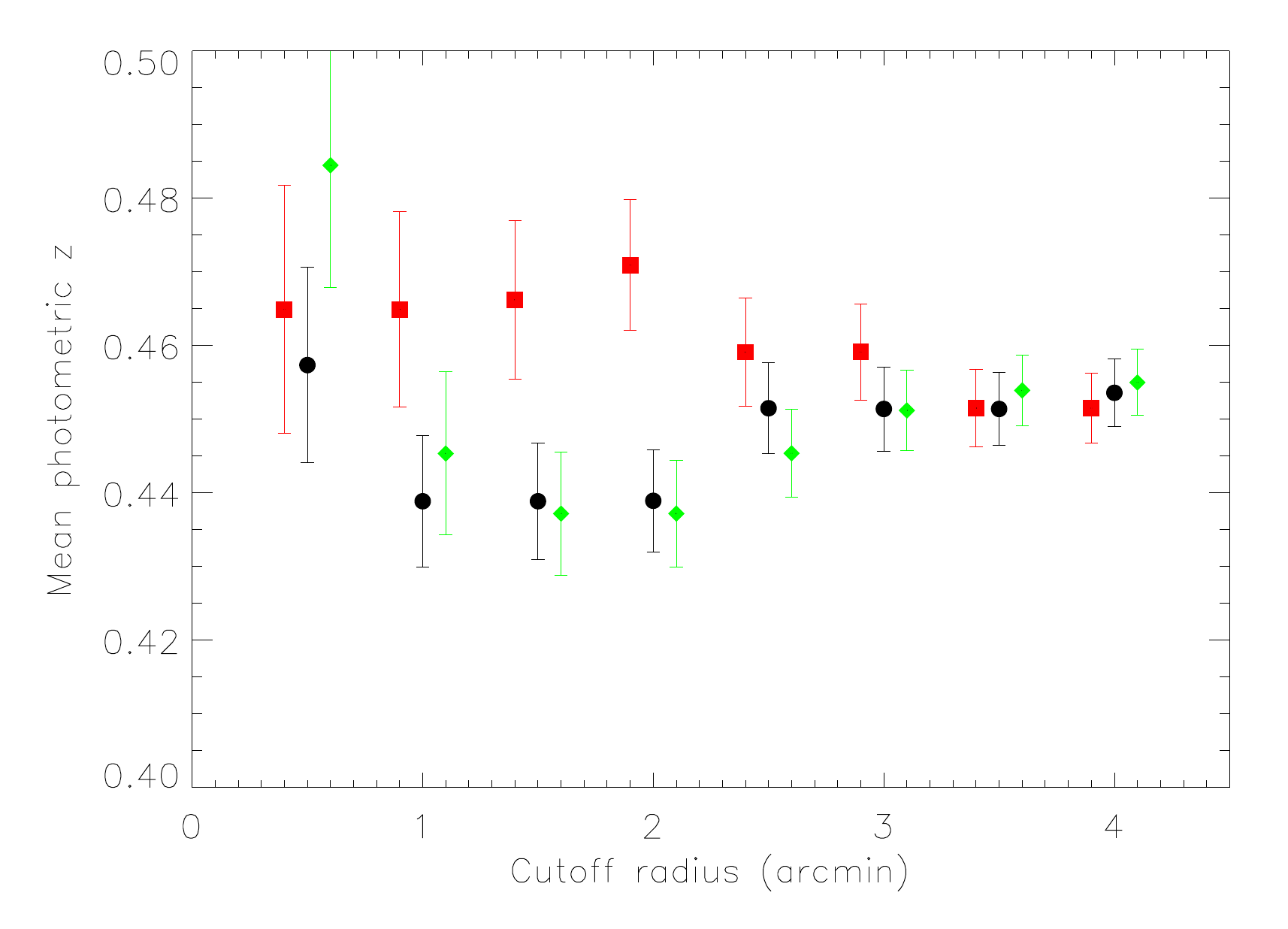}
\caption{Median photometric redshifts for the three clumps ($A$ black
circles, $B$ red squares, $C$ green diamonds) of all the galaxies
within the cutoff radius. The error bars are the standard deviations
of the photometric redshifts distribution. Given the small separation between
components $A$ and $B$, the points at radii $\apgt 2\arcmin$ may be
contaminated by galaxies of the other structure.}
\label{fig:photoz}
\end{center}
\end{figure}

We used the archival photometric redshifts in our catalogue to
estimate the redshift of the three components.  
We extracted a sub-catalogue selecting only galaxies in the
photometric redshift range $0.35-0.6$ and, for each clump, we 
calculated the median redshift of the galaxy population around
the X-ray centre as a function of the cutoff radius. The resulting
plot is shown in Fig.\,\ref{fig:photoz}.
Components $A$ and $C$ are both consistent with the spectroscopic value of
their central galaxies ($z=0.45$), but
the innermost $2 \arcmin$ of component $B$
indicate a slightly larger redshift ($z_{phot} \simeq 0.47$), similar to the
results of the X-ray analysis, although consistent at two $\sigma$ with the
values of the other components. At
larger radii, the redshift estimates for the three clumps are all consistent
with each other. It should be noted however that components
  $A$ and $B$ are separated only by $\simeq 2$ arcmin, thus all the
  estimates at similar or larger radii may be contaminated by galaxies
  belonging to the other cluster.

\subsection{Optical appearance and morphology of the cluster}
\begin{figure*}
\begin{center}
\includegraphics[keepaspectratio,width=\textwidth]{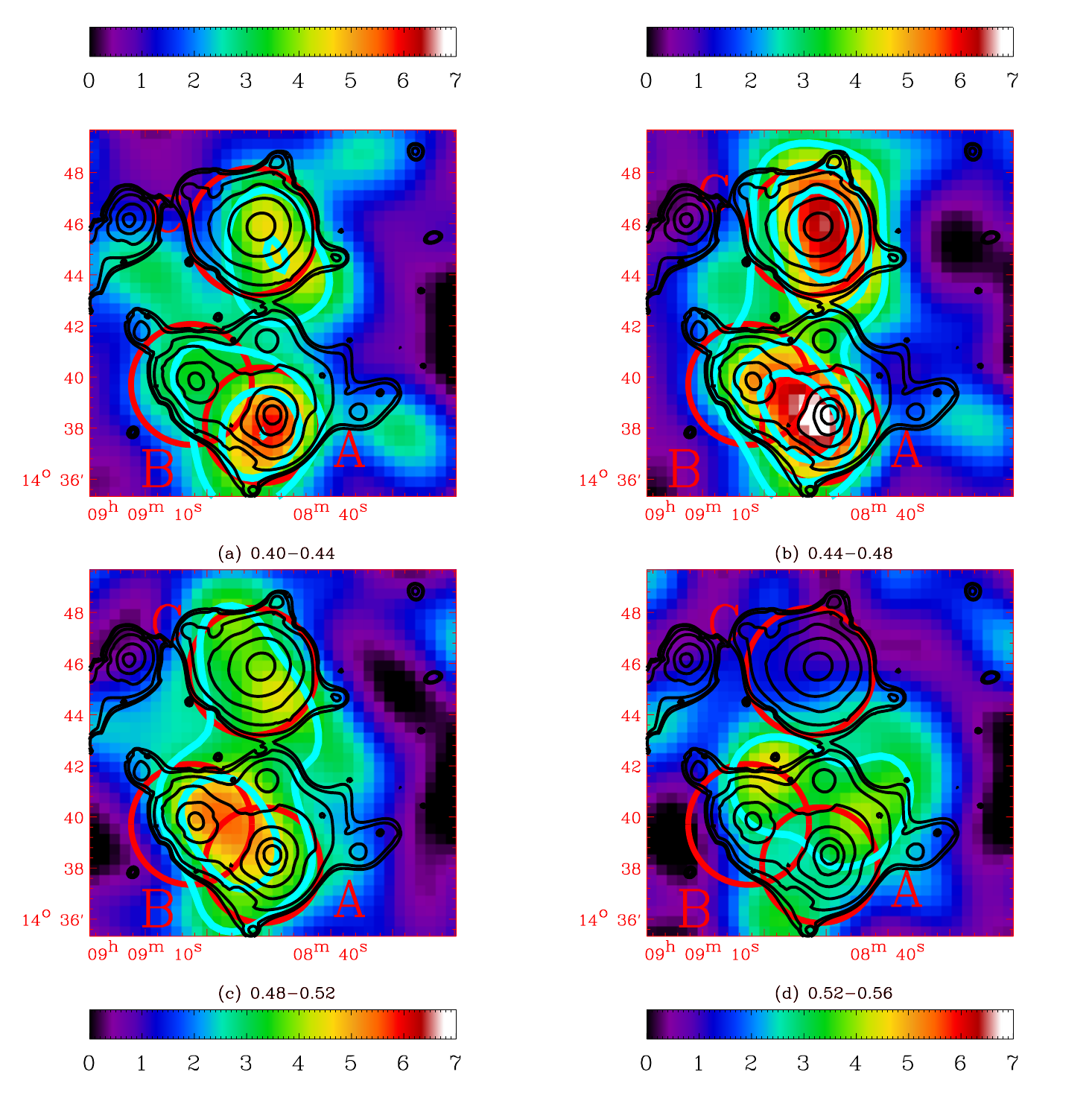}
\caption{Galaxy density maps for cluster members in the SDSS catalogue
(colours) in different photometric redshift cuts: $0.40-0.44$ (upper left),
$0.44-0.48$ (upper right), $0.48-0.52$ (lower left) and $0.52-0.56$
(lower right). The cyan
contours overlaid mark the significance of each density peak at 3, 4, and 5
$\sigma$, respectively. The black contours show the X-ray distribution
and the red circles and letters mark the three components.}
\label{fig:galdens_zcut}
\end{center}
\end{figure*}

We used the catalogue from SDSS to build two-dimensional
galaxy density maps in different photometric redshift cuts (Fig.
\ref{fig:galdens_zcut}), with a width $\Delta z_{phot}=0.04$. 
We assigned the galaxies to a fine grid
of $24\arcs$ per pixel, which is then degraded with a
gaussian beam to an effective resolution of 3 arcminutes. We also computed a
significance map using as reference ten random non overlapping control regions
in a 9 deg$^2$ area around the system.\\
Clear galaxy overdensities show up around $z_{phot}=0.46$
at the location of the three X--ray clumps. %
However, these overdensities do not appear isolated. At
the
location of cluster $B$, we see an overpopulation of galaxies towards higher
redshift ($5\sigma$ peak at $z_{phot}\sim0.5$), consistent with the
redshift $z=0.48$ we found in X-rays,
whereas the overdensity
extends towards slightly lower redshifts ($z_{phot}\sim0.42$) at the position of
cluster
$A$. There are also indications of another concentration close to component $B$
at larger redshift ($0.52-0.6$).\\
We investigated the maps in Fig. \ref{fig:galdens_zcut} to look for a possible
population of inter-cluster galaxies: i.\,e.\,, objects not associated with one of the
three clumps but rather with the whole structure, which would support a
scenario where the three clumps are physically connected. 
We draw iso-contours levels in the significance map (Fig.\,
\ref{fig:galdens_zcut}): the outermost contour between $0.44$ and
$0.52$ connecting the three clumps indicates the presence of a
3$\sigma$ excess in the galaxy number density above background in the
inter-cluster region.
%and indeed we
%found, between $0.44$ and $0.52$,  a connection between the three clumps,
%significant at a three $\sigma$ confidence level. \\
%LK:  it is not clear to me what this means to have a "connection... significant at a three $\sigma$ confidence level." Do you mean that everywhere between the galaxy clusters, the number densities of galaxies is more than 3 sigma above background?

\section{Comparison with \Planck}

\subsection{Total SZ signal}
As a simple comparison of the SZ and X-ray properties, we can compare the
\planck\ $Y$ measurement with
 the predicted values from the sum of the
$\YX$ of all three components, using the scaling relations in
\citet{arnaud10}. From our X-ray estimates (Table 1), we predict
the total integrated value of the Comptonization parameter
within a sphere of radius $5R_{500}$ for the sum of the three components to be
$Y_{x,5R500}=(7.52 \pm
0.9)\cdot 10^{-4}$ arcmin$^{2}$. This is about $50\%$ of the measured signal in
the same region which was found in Paper I and the two values are compatible at
$2.3\,\sigma$. In the following, we will work under the
  assumption that the three clusters are all located at the same
  redshift $z=0.45$. Considering the best fit redshift
  for component $B$ leads to a  slightly smaller SZ flux   $Y_{x,5R500}=(6.44 \pm
0.7)\cdot 10^{-4}$ arcmin$^{2}$.  
As discussed in Sec. \ref{s:ob:planck}, we used the parameters provided in Table
\ref{table:x_prop} to improve our estimate of the total SZ signal of this
structure from \Planck\ data. 
We built a specific template from the X-ray
analysis, made from three universal pressure profiles cut to $5\R500$
\citep{arnaud10}
corresponding to the three components. Each component is placed at its precise
coordinates and the size is given by the $\R500$ value in Table
\ref{table:x_prop}.
We also fixed the relative intensity between the components to verify
A/C$=0.96/1.61$ and B/C$=1.22/1.61$ for the ratio of integrated
 fluxes. Then we ran the MMF3 algorithm \citep{mel06}  to estimate the amplitude
of the template three times, centering the maps on components $A$, $B$ and $C$.
The MMF3 algorithm estimates the noise (instrumental and astrophysical) in
a region of $10\times10$ deg$^2$  around the centre (excluding
  the region within $5\R500$), therefore
changing the centering from one component to the
other can affect the background estimation and therefore the flux and
signal-to-noise ratio. Centering maps on component $A$ we found
$Y_{5\R500}=(9.75
\pm
3.19)\,10^{-4}\rm{arcmin}^2$, on $B$ $Y_{5\R500}=(12.26 \pm
3.22)\,10^{-4}\rm{arcmin}^2$ and on $C$ $Y_{5\R500}=(12.97 \pm
3.20)\,10^{-4}\rm{arcmin}^2$.
Our SZ flux estimations are all compatible with each other. They
are also slightly larger than
the X-ray prediction but consistent at $0.7 - 1.7\, \sigma$
($0.9-1.9\,\sigma$ using $z=0.516$ for component $B$). We further allowed
the position of the template to be a free parameter and found that the algorithm
is able to reconstruct the position of the peak with a positional accuracy of
one sky pixel ($1.717\times 1.717<$ arcmin$^2$, in a HEALPIX projection of
$nside=2048$, \citealt{gorski2005}). 
This is consistent with the positional accuracy of the MMF3 algorithm, which
has been tested both on simulations and on real data with known
clusters. \\
The discrepancy between the observed SZ signal and the prediction from the $Y_X$
measurement is decreased with respect to Paper I:
while the X-ray prediction was only 40\% of the SZ measurement, it is now 
between 60 and 77\%, depending on the map centering.
This is partly due to the
larger $\YX$ values we found in this analysis with respect to Paper I,
especially for components $B$ and $C$. It is also certainly due to the improved
accuracy of the HFI maps obtained with two full surveys of the sky and to the
multi-component model we have used to estimate the SZ flux, with respect to the
data from the first sky survey and to the single component model that was used
in Paper I. Indeed, these results
confirm our capability to extract faint SZ signals,
when guided by X-ray priors \citep{planck2011-5.2a}. \\

\subsection{SZ signal distribution}

It is possible to combine the X-ray images with the temperatures of the
components to predict the distribution of the SZ signal (see
\citealt{mroc12} for a similar approach). X-ray images in the
soft band are proportional to the square of the density integrated along the
line of sight and therefore their square root can be combined with a
temperature map to derive a pseudo-pressure map\footnote{Although
  deriving pseudo-pressure maps as discussed in the text is customary
  in the literature, we underline here that this approach is not
  completely valid. The X-ray surface brightness, ignoring the
  temperature dependance, is proportional to $\int n^2dl$ and its
  square root is never equal to $\int n dl$, which is the expression
  that should enter in the definition of the comptonization parameter
  $y$. However, pseudo-pressure maps can still be used for qualitative
  comparison with the $y-$maps.}, which when smoothed with the
\Planck\ resolution, can be qualitatively compared with the $y-$maps.
 We combined the background subtracted X-ray image with a temperature map,
built assuming 
the mean temperature value 
in each component (Table \ref{table:x_prop}) within $\R500$ and zero outside, to
produce a pseudo-pressure map,
that we smoothed with a gaussian filter of 10\arcmin\ FWHM to mimic the
resolution of \planck\ $y-$maps.
\Planck\ cannot spatially resolve the three components of this object, therefore
we expect the peak of the pseudo-pressure map to be located around the
barycentre of the system just for resolution effects. 
\begin{figure*}
\begin{center}
\includegraphics[keepaspectratio,width=\textwidth]{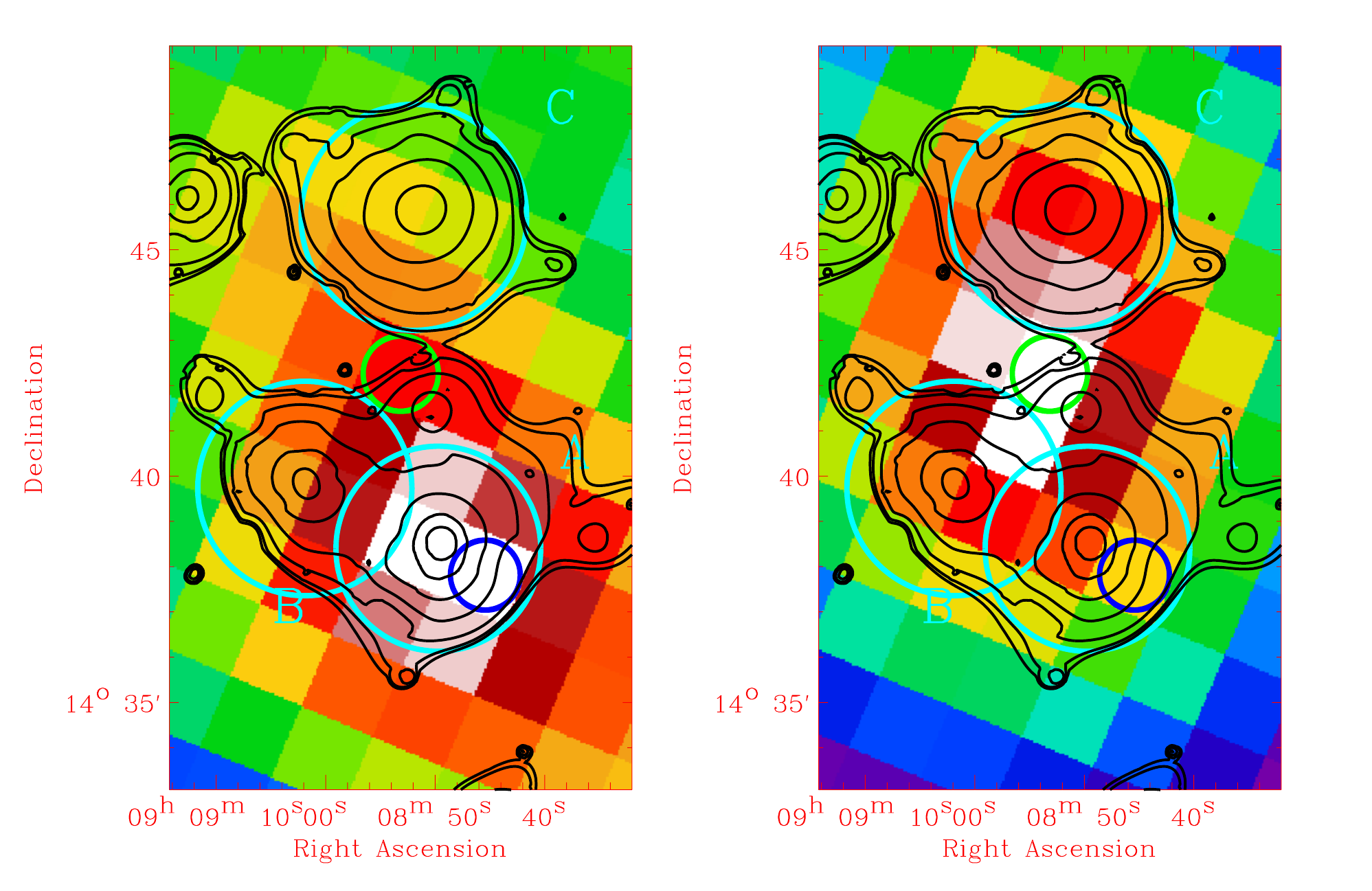}
\caption{MILCA $y-$map (left panel) compared with the pseudo pressure X-ray map
degraded to the $y-$map resolution (right panel). X-ray contours and
cyan circles indicating the three components are over-plotted
to guide the eye. The green and blue circles mark the positions of the peaks in
the $y-$map and in the
pseudo-pressure
distribution, respectively.}
\label{fig:cfr_map}
\end{center}
%\label{fig:cfr_sbprof}
\end{figure*} 
The results are shown in Fig. \ref{fig:cfr_map}, compared with the MILCA $y-$map.
The position of the peak in the SZ map does not coincide with the peak of the
pseudo pressure map: while the latter is located 
as expected at the barycentre between the three components, the $y-$map suggests
an excess of pressure to the SW of component $A$. The offset between the two
peaks is $\simeq 5$\arcmin.\\ 
We have performed some tests both on the X-ray and on the SZ maps to
investigate the origin of this offset. On the X-ray side, we
have
produced surface brightness images using a different background modelling. The 
first test concerned the background subtraction: we
used the ESAS
software\footnote{\url{http://heasarc.nasa.gov/docs/xmm/xmmhp_xmmesas.html}} to
produce particle background and residual soft proton images and we created
images of the ''sky background'' components (CXB and galactic foregrounds) 
modelling them in an external annulus \citep{lec08} and rescaling them across
the field of view \citep{ett10}. Point sources could also affect the position
of the pseudo pressure peak, therefore we ran a different point source
algorithm using the SAS task {\verb edetect_chain } on \mos\ and \pn\ images in
five energy bands and we added undetected sources we identified
with a visual inspection of the images. Both these tests showed a negligible
impact on the position of the peak, which in fact is located where it is
expected to be, at the barycentre of the three components. \\
For the SZ effect, we have compared the maps reconstructed with different 
ILC-based algorithms. Besides MILCA,
we tested  GMCA \citep{bob08} and NILC \citep{del09} algorithms (see Planck
intermediate
paper: SZ and pressure profile of galaxy clusters, Planck collaboration
2012, for a summary description and a comparison at the cluster scale of the
three methods). The three $y-$maps are very consistent and the position of the
peak does not change across the
maps. \\
Furthermore, we noticed that the offset was not found in the previous
version of the maps, which was shown in Paper I. \\
The presence of correlated noise in the {\it y}-map produced from 
\planck\ data can be a major source of error in the reconstruction of the 
position of clusters, and in particular in the case of low signal to 
noise systems such as \name. To quantify this error we have 
produced Monte Carlo simulations. We first estimated the noise covariance 
matrix on the \planck\ Comptonization parameter map of this system and
we produced 500 realisations of noise to each of 
which we added the expected thermal SZ effect from \name\ to construct 
mock $y-$maps.
Then, we estimated the position of the supercluster in each of these maps 
accounting for correlation of the noise. Finally, we computed the average 
and the standard deviation of the error on the reconstructed position and
we obtained an average error of $(4.5 \pm 2.5) \arcmin$.  Therefore, the offset
between the reconstructed positions from the \planck\
$y-$map and the pseudo-pressure X-ray derived map of \name\ is consistent with
being due
to noise. The same applies for the separation between the peak
in the \planck\ $y-$ map we show here and the one which was shown in Paper
I.\\

\section{Discussion and conclusion}
The first observations of the multi-wavelength follow-up campaign of \name, a
triple system of galaxy clusters discovered by \planck, have allowed us to
improve our understanding of this object. With the new \xmm\ observation
we estimated the global properties of each component: the ICM temperatures range
from $3.5$ to 5 keV and the total masses within $\R500$ are in the range
$2.2-3\, 10^{14} \msol$. We detected the iron K$\alpha$ lines in the X-ray
spectra of
each component, and therefore we were able to confirm that components $A$ and $C$
are lying at the same redshift ($z=0.45$). However, given the large angular
separation of these two components ($7.5 \arcmin$, corresponding to $2.6$ Mpc,
in the plane of the sky), they have likely not started to interact yet and we
did
not detect significant excess X-ray emission between these two components.
For component $B$, we
estimated a larger redshift from  X-ray spectroscopy ($z=0.48$), although
consistent at  two $\sigma$ with the best fit value for component $A$.
A similar indication is supported by the optical data, with the photometric
redshifts we retrieved from SDSS DR8. 
However, given the large uncertainties
of our redshift estimates (based both on X-rays and on photometry), a more
detailed
picture of the three-dimensional structure of \name\ 
will be possible only with the measurement of 
spectroscopic redshifts for a large sample of member galaxies, that is already
foreseen with VLT in our follow-up program.\\
Our redshift results are consistent with the three clusters being part of the
same supercluster structure, that will eventually lead to the formation of a 
massive object ($\simeq 10^{15}\msol$). This is supported also by our analysis
of the galaxy population with SDSS data: the
galaxy density maps show the presence of a possible population of inter-cluster
galaxies, significant at $3\sigma$, connecting the whole system (Fig.\,
\ref{fig:galdens_zcut}). 
However, the relaxed appearance of
component $A$, its large distance ($2.5$ Mpc) in the plane of the sky from
component $C$ and along the line of
sight from component $B$, as well as the absence of any detectable excess X-ray
emission between the components may suggest that we are witnessing a very early
phase of interaction.\\
Using the X-ray results from the new \xmm\ observation, we built a
multicomponent model that we used to extract the total SZ signal from \Planck\
data. We compared the improved
estimate of $Y_{SZ}$ with the prediction from X-rays and we found the
latter to be about 68\% of the measured SZ signal. The discrepancy
between these two values is reduced with respect to Paper I and is
only marginally significant at $1.2\sigma$.  \\
The results from our simulations have shown that an offset as large as
$5\arcmin$ can be expected in the reconstructed $y-$maps for low significance
objects, due to noise fluctuations
and astrophysical contributions. With this study we have illustrated the
expected difficulty of accurately reconstructing the two-dimensional SZ signal for
objects with low signal-to-noise ratio. Indeed the instrumental noise and
astrophysical contamination compete seriously with the SZ effect at the
detection
limit threshold. Nonetheless, objects like \name\ can be detected with a
dedicated optimal filtering detection method, and the SZ signal can be
reconstructed assuming priors (such as position, size and relative intensity)
from other wavelengths.\\
 Despite a deep re-observation of this system with \xmm,
the intrinsic limitations of our X-ray data and of the current \planck\ SZ maps
do not allow us for the time being to assess the presence of possible
inter-cluster emission.\\
A careful
analysis of the galaxy dynamics in the complex potential of this object and of
the mass distribution from weak lensing will both be available with our
on-going optical follow up program. These observations, combined with the
results presented in this paper and with new \planck\ data obtained in two
other full surveys of the sky, might deliver further clues for the
understanding of this peculiar triple system.

%
%________________________________________________________________
\begin{acknowledgements}
A description of the Planck Collaboration and a list of its members,
indicating which technical or scientific activities they have been
involved in, can be found at \url{http://www.rssd.esa.int/Planck}. The
Planck Collaboration acknowledges the support of: ESA; CNES and
CNRS/INSU-IN2P3-INP (France); ASI, CNR, and INAF (Italy); NASA and DoE
(USA); STFC and UKSA (UK); CSIC, MICINN and JA (Spain); Tekes, AoF and
CSC (Finland); DLR and MPG (Germany); CSA (Canada); DTU Space
(Denmark); SER/SSO (Switzerland); RCN (Norway); SFI (Ireland);
FCT/MCTES (Portugal); and DEISA (EU). The present paper is also partly
based on observations obtained with \xmm\, an ESA science mission with
instruments and contributions directly funded by ESA Member States and
the USA (NASA), and on data retrieved from SDSS-III. Funding for SDSS-III has been provided by the Alfred P. Sloan Foundation, the Participating Institutions, the National Science Foundation, and the U.S. Department of Energy Office of Science. The SDSS-III web site is http://www.sdss3.org/.
SDSS-III is managed by the Astrophysical Research Consortium for the Participating Institutions of the SDSS-III Collaboration including the University of Arizona, the Brazilian Participation Group, Brookhaven National Laboratory, University of Cambridge, Carnegie Mellon University, University of Florida, the French Participation Group, the German Participation Group, Harvard University, the Instituto de Astrofisica de Canarias, the Michigan State/Notre Dame/JINA Participation Group, Johns Hopkins University, Lawrence Berkeley National Laboratory, Max Planck Institute for Astrophysics, Max Planck Institute for Extraterrestrial Physics, New Mexico State University, New York University, Ohio State University, Pennsylvania State University, University of Portsmouth, Princeton University, the Spanish Participation Group, University of Tokyo, University of Utah, Vanderbilt University, University of Virginia, University of Washington, and Yale University. 

\end{acknowledgements}

\bibliographystyle{aa}
\bibliography{sc,Planck_bib}

\end{document}